\documentclass[11pt]{article}

\usepackage[pagewise]{lineno}
\usepackage[utf8]{inputenc}
\usepackage{graphicx}
\graphicspath{{figs/}}
\usepackage{subcaption}
\usepackage{amsmath, amssymb}
\numberwithin{equation}{section}
\usepackage{xcolor}
\usepackage{natbib}
\usepackage{bm}
\usepackage{nicefrac}
\usepackage{booktabs}
\usepackage{tabularx,tabulary,array}
\newcolumntype{M}[1]{>{\centering\arraybackslash}m{#1}}
\newcolumntype{N}{@{}m{0pt}@{}}
\usepackage{multirow}
\usepackage{dcolumn}
\usepackage{lscape}
\usepackage[normalem]{ulem}
\usepackage[title]{appendix}
\usepackage{authblk}
\usepackage{hyperref}

\usepackage{graphicx}
\usepackage{subcaption}
\usepackage{mwe}

\usepackage{authblk}
\usepackage[english]{babel}
\usepackage[T1]{fontenc}

\usepackage{amssymb, amsmath, amsthm, blkarray}
\usepackage{graphicx, tikz}
\usepackage{natbib}

\definecolor{mylightyellow}{rgb}{1,1,.8}
\definecolor{mylightgreen}{rgb}{.8,1,.8}
\definecolor{mydarkred}{RGB}{178,34,34}
\definecolor{mydarkgreen}{RGB}{34,139,34}
\definecolor{mydarkblue}{RGB}{72,61,139}
\definecolor{mydarkyellow}{RGB}{218,165,32}

\usepackage{booktabs}

\usepackage{marginnote}
\usepackage{soul}

\usepackage[a4paper,margin=25mm]{geometry}

\usepackage[]{hyperref}
 \hypersetup{
 colorlinks=true,breaklinks=true,
 urlcolor= mydarkblue,linkcolor= mydarkblue,citecolor= mydarkgreen,
 pdfauthor={S\'aez de Oc\'ariz Borde, H., Sondak, D., Protopapas, P.}
}

\theoremstyle{plain}
\theoremstyle{definition}

\usepackage{eurosym}

\title{Convolutional Neural Network Models and Interpretability for the Anisotropic Reynolds Stress Tensor in Turbulent One-dimensional Flows}

\author[1]{Haitz S\'aez de Oc\'ariz Borde\thanks{\tt haitz.saez-de-ocariz-borde17@imperial.ac.uk}}
\author[2]{David Sondak\thanks{\tt dsondak@seas.harvard.edu}}
\author[2]{Pavlos Protopapas\thanks{\tt pavlos@seas.harvard.edu}}

\affil[1]{\small Department of Aeronautics, Imperial College London, London SW7 2AZ, United Kingdom}
\affil[2]{\small Institute for Applied Computational Science, Harvard University, Cambridge, MA 02138,
United States}

\date{
\small First Version: March 11, 2021.  This version: \today
}

\begin{document}

\maketitle

\begin{abstract}

The Reynolds-averaged Navier-Stokes (RANS) equations are widely used in turbulence applications. They require accurately modeling the anisotropic Reynolds stress tensor, for which  traditional Reynolds stress closure models only yield reliable results in some flow configurations. In the last few years, there has been a surge of work aiming at using data-driven approaches to tackle this problem. The majority of previous work has focused on the development of fully-connected networks for modeling the anisotropic Reynolds stress tensor. In this paper, we expand upon recent work for turbulent channel flow and develop new convolutional neural network (CNN) models that are able to accurately predict the normalized anisotropic Reynolds stress tensor. We apply the new CNN model to a number of one-dimensional turbulent flows. Additionally, we present interpretability techniques that help drive the model design and provide guidance on the model behavior in relation to the underlying physics.

\end{abstract}

\bigskip

\noindent {\bf Keywords:} turbulence modeling, Reynolds-averaged Navier-Stokes, deep learning, convolutional neural networks, interpretability.

\newpage



\section{Introduction}
\label{sec:introduction}

Most real-world fluid problems and applications involve turbulent flow. As such, engineers and scientists remain concerned about developing reliable turbulence models that have good performance and that at the same time are practical and tractable from a computational perspective. Given that direct numerical simulations (DNS) of turbulent flow are too computationally expensive at the Reynolds numbers of interest, researchers have developed a number of alternative simplified and more practical turbulence models. Two popular approaches are Reynolds-averaged Navier-Stokes (RANS) models and large eddy simulation (LES) models. RANS models are less computationally expensive than LES models, but also less accurate. RANS models aim to determine the average velocity and pressure fields and to model the effects of the fluctuating components on the average. The effect of the fluctuating fields arises due to the  Reynolds stress tensor, which must be modelled. Although more traditional two-equation eddy viscosity models such as the $k-\epsilon$ and the $k-\omega$ are commonly used in many engineering applications, their applicability to different flows is limited (see~\cite{speziale,johansson2002engineering, chen2003extended, gatski2004constitutive}). 

Recently, several researchers have worked on applying machine learning algorithms to develop turbulence closure models for RANS, such as~\cite{ling_ml}, \cite{ling2016reynolds}, \cite{fang2018deep}, \cite{PhysRevFluids.3.074602}, \cite{Kaandorp2018MachineLF}, \cite{song}, \cite{KAANDORP2020104497} and~\cite{zhu2020datadriven}. \cite{ling2016reynolds} proposed the Tensor Basis Neural Network (TBNN) to learn the coefficients of an integrity basis for the Reynolds anisotropy tensor from~\cite{pope_1975}. The TBNN automatically
guarantees the physical invariance in the predicted Reynolds anisotropy tensor. Inspired by \cite{ling2016reynolds}, \cite{fang2018deep} used machine learning to model the Reynolds stress for fully-developed turbulent
channel flow. They used a generic fully-connected feedforward neural network (FCFF) as the base model and increased its capabilities by using non-local
features, directly incorporating friction Reynolds number information, and enforcing the boundary condition at the channel
wall. A survey regarding recent work in data-driven turbulence modeling can be found in~\cite{duraisamy}.

In this paper, we present a convolutional neural network (CNN) architecture that outperforms the fully-connected architectures developed by~\cite{fang2018deep} for the one-dimensional turbulent channel flow problem. We train and test our new architecture on one-dimensional turbulent flow configurations and demonstrate that the new neural network works beyond the simple channel flow. We also propose and conduct several interpretability analyses to identify significant regions of the flow and guide the architecture design.

Although an exact definition for interpretability has not been developed yet, it could be described as the ability to explain in understandable terms the choices made by a machine learning algorithm to a human. Interpretability has recently gathered a lot of attention in the machine learning community. See~\cite{zhang2021survey} for a survey on some of the latest interpretability techniques. Other less recent surveys on interpretability are those by~\cite{electronics8080832} and by~\cite{8397411}. There is a growing interest in understanding how machine learning models make decisions and how these relate to the underlying problem. It is unlikely that machine learning augmented turbulence models will be ultimately successful without the need for direct human expertise and hence, understanding the algorithm decision making process is key to improving the models. Most of the work on interpretability so far has been quite general, such as that by~\cite{molnar2019},~\cite{gilpin2019explaining} and~\cite{doshivelez2017rigorous}. These works serve as a good introduction to interpretability, but only present generic algorithms that do not directly address interpretability in a turbulence modelling context. To the best of our knowledge, interpretability techniques have not been applied to machine learning in turbulence modeling to guide architecture design. In this paper, we will try to address this issue and adapt some interpretability techniques used in other machine learning applications to our turbulence modelling problem.

The rest of the paper has the following structure. Section \ref{sec:Background} covers background information on the RANS equations, several one-dimensional turbulent flows, the datasets used, and neural networks and interpretability. Section \ref{sec:Neural Network Models} reviews the neural network models by~\cite{fang2018deep} and introduces our new convolutional models. Section \ref{sec:Interpretability Techniques} describes the interpretability techniques that have been adapted for our turbulence modelling problem. Section \ref{sec:Results} covers the performance of the different machine learning models and presents the interpretability results. Lastly, Section \ref{sec:Conclusion} summarizes the final conclusions and suggests future work directions.

\section{Background and Methodology}
\label{sec:Background}

The present work aims at developing CNN models to represent the Reynolds anisotropy tensor for turbulent one-dimensional flows, as well as at applying interpretability techniques to understand how the models make decisions and how these relate to the underlying physics of the problem. In this section, some background on turbulent channel flow, turbulent Couette flow, and turbulent channel flow with wall transpiration, neural networks, and interpretability are provided.

\subsection{The RANS Equations}
\label{subsec:The RANS equations}

The Reynolds-averaged approach decomposes the velocity into average and fluctuating components $\mathbf{u}=\overline{\mathbf{u}}+\mathbf{u^{\prime} }$ to find the average field $\overline{\mathbf{u}}$. This averaging operation applied on the Navier-Stokes equations leads to the RANS equations. The RANS equations include an additional term $\overline{\mathbf{u^{\prime}}\otimes\mathbf{u^{\prime}}}$ known as the Reynolds stress tensor, and this must be modelled to close the RANS equations. A substantial portion of modelling efforts have concentrated on the anisotropic Reynolds stress tensor $\mathbf{a}=\overline{\mathbf{u^{\prime}}\otimes\mathbf{u^{\prime}}}-(2k/3)\mathbf{I}$ as that is the portion responsible for turbulent transport, where $k$ is the turbulent kinetic energy $k=\frac{1}{2}\text{trace} (\overline{\mathbf{u^{\prime}}\otimes\mathbf{u^{\prime}}})$. In this work, the models are trained on the normalized anisotropy tensor, $\mathbf{b} = \mathbf{a}/(2k)$, and individual components of the tensor are denoted with subscripts referring to the fluctuating velocity components involved (e.g. $b_{uv} = \overline{u^{\prime}v^{\prime}} / (2k)$).

Eddy viscosity models such as the $k-\epsilon$ and the $k-\omega$ models are some of the most popular models used to obtain predictions for the anisotropic Reynolds stress tensor, but they do not provide satisfactory results for all flows (see e.g.~\cite{speziale,johansson2002engineering, gatski2004constitutive, chen2003extended}). Recently, machine learning algorithms have been proposed to tackle this problem and obtain predictions for the anisotropic Reynolds stress tensor in a variety of flows, including those for which eddy viscosity models are inadequate (see e.g. \cite{Tracey2015AML, Zhang2015MachineLM, ling2016reynolds, PhysRevFluids.3.074602, wang}). In particular, the TBNN by~\cite{ling2016reynolds} has provided improved predicative accuracy compared to
traditional RANS models when trained and tested on several flows.

\subsection{Turbulent Channel Flow}
\label{subsec:Turbulent channel flow}
Turbulent channel flow is a paradigmatic flow in which a fluid confined between two parallel plates in the $x-z$ plane is driven by a pressure gradient in the streamwise direction. The streamwise, spanwise, and wall-normal directions are denoted by $x$, $z$, and $y$, respectively.
The goal is to determine the velocity field $\mathbf{u}=(u,v,w)$, where $u$, $v$, and $w$ are the velocities in the $x$, $y$, and $z$ directions. In general, each component of the velocity field depends on space $(x,y,z)$ and time $t$. The friction velocity $u_{\tau}=\sqrt{\tau_{\text{wall}}/\rho}$ is a natural velocity scale for this problem, where $\tau_{\text{wall}}$ is the wall shear stress and $\rho$ is the fluid density. Selecting the channel half-width $h$ and the friction velocity as the length and velocity scales of the problem leads to the friction Reynolds number $Re_{\tau} = u_{\tau}h/\nu$ as a natural Reynolds number for this flow, where $\nu$ is the kinematic viscosity of the fluid. Results presented in this paper are typically shown in wall-units, $y^{+} = u_{\tau}y/\nu$. Turbulent channel flow is statistically one-dimensional, its mean velocity field being $\overline{\mathbf{u}}=(\overline{u}(y),0,0)$. In this case, the axial momentum equation is solved for $\overline{u}\left(y\right)$ and the only component of $\mathbf{a}$ affecting the velocity profile is $a_{uv}$. 


The machine learning models in this work are trained and tested on the same turbulent channel flow dataset used by~\cite{fang2018deep}, which consists of DNS data at four friction Reynolds numbers $Re_{\tau} = [550, 1000, 2000, 5200]$. The number of DNS data points varies for different
friction Reynolds numbers due to differing mesh resolution requirements. For $Re_{\tau}=[550, 1000, 2000, 5200]$ the dataset has $N_{y}=[192, 256, 384, 768]$ respective points in the wall-normal direction. The data is accessible at the Oden Institute turbulence file server\footnote{https://turbulence.oden.utexas.edu/}. Detailed information on the DNS simulations can be found in~\cite{lee_moser_2015}. 



\subsection{Turbulent Couette Flow}
\label{subsec:Turbulent Couette flow}

Turbulent Couette flow is a shear-driven fluid flow in which the fluid is confined between two parallel surfaces one of which is moving tangentially relative to the other. This flow configuration is a proxy for some practical problems, namely, the Earth's mantle movement or extrusion in manufacturing processes. In our case, we consider the flow between two parallel planes with no pressure gradient. Similarly to turbulent channel flow, our turbulent Couette flow configuration is also statistically one-dimensional with a mean velocity field $\overline{\mathbf{u}}=(\overline{u}(y),0,0)$, and to find the solutions to the RANS equations $\overline{u}$ must be computed, which depends on $a_{uv}$. For this flow, the friction Reynolds number $Re_{\tau}$ is defined based on the friction velocity and the channel half-width, as suggested by~\cite{dns_couette}, to remain consistent with other studies regarding channel flow, such as those by~\cite{lduran} and~\cite{lee_moser_2015}. The motivation for using this definition of $Re_{\tau}$ is to ease comparison between dissimilar flows. Nevertheless,~\cite{barkley_tuckerman_2007} suggested that the appropriate length scale in the case of Couette flow should correspond to the full width according to their study of transitional flows. 

The study of planar Couette flow has been more limited in total number of studies and the variety of Reynolds numbers investigated compared to channel flow, both from a experimental and computational perspective. \cite{dns_couette} attributed this fact to the existence of very-large-scale motions in Couette flow, which is hard to represent in DNS simulations, and tried to address this problem in their paper. We used the DNS data by~\cite{dns_couette} to train and test our model in Couette Flow. The data consisted of three friction Reynolds numbers $Re_{\tau} = [93, 220, 500]$ with $N_{y}=[64,96,128]$ points in the wall-normal direction. The data for turbulent Couette flow is also accessible at the Oden Institute turbulence file server.

\subsection{Turbulent Channel Flow with Wall Transpiration}
\label{subsec:Turbulent Channel Flow with Wall Transpiration}

In addition to the two paradigmatic flows discussed above, we also trained and tested our model on fully-developed turbulent channel flow with wall transpiration. In this case, uniform blowing and suction was used on the lower and upper walls, respectively. It is well known that in the classical plane turbulent channel flow without transpiration, all statistical quantities of the flow are either symmetric or antisymmetric with respect to the channel centerline. However, when wall transpiration is included, these symmetries do not hold, which modifies the scaling laws for the linear viscous sublayer and the law of the wall for wall-bounded flows with no transpiration as discussed by~\cite{wall_tr}. This makes turbulent channel flow with wall transpiration an interesting flow case to train and test our model, which will need to be able to capture substantially more complicated flow features than those for channel and Couette flow. For this flow, the friction velocity $u_{\tau}$ is defined as the mean friction velocity $u_{\tau}=\sqrt{\frac{u_{\tau b}^{2}+u_{\tau s}^{2}}{2}}$, where subscripts $b$ and $s$ denote variables on the blowing
and the suction side. Once more, in the case of turbulent channel flow with wall transpiration, $\overline{u}$ must be found to solve the RANS equations, which is only dependent on $a_{uv}$.

The DNS data used for turbulent channel flow with wall transpiration had three friction Reynolds numbers $Re_{\tau} = [250, 480, 850]$ with $N_{y}=[251,385,471]$ respective points in the wall-normal direction. The data is available online at the TU Darmstadt Fluid Dynamics Turbulence DNS data base\footnote{https://www.fdy.tu-darmstadt.de/fdyresearch/dns/direkte$\_$numerische$\_$simulation.en.jsp}. Note that the data is provided at a number of $\frac{V_0}{u_{\tau}}$ ratios, where $V_0$ refers to the transpiration velocity. In our case, we use the data for $\frac{V_0}{u_{\tau}}=5$. For the DNS simulation, \cite{wall_tr} implemented a numerical code developed at the School of Aeronautics, Technical University of Madrid, which is described in detail by~\cite{hoyas}. 

\subsection{Neural Networks}
\label{subsec:Neural Networks}
Deep feedforward networks, also called feedforward neural networks or multilayer perceptrons (MLPs), are considered the archetype of deep learning (DL) models. Feedforward networks are used to obtain an approximation of some function $f$. For instance, $\mathbf{y}=f(\mathbf{x})$ maps an input $\mathbf{x}$ to a prediction $\mathbf{y}$. A feedforward network defines a mapping $\mathbf{y}=f^{*}(\mathbf{x};\boldsymbol{\theta})$ and is optimized to find the parameters $\boldsymbol{\theta}$ that give the best approximation of the function $f$. In feedforward models the information flows through the function being evaluated from $\mathbf{x}$, through the intermediate computations, and finally to the output $\mathbf{y}$, as described by~\cite{Goodfellow-et-al-2016}. 

Neural networks can usually be represented by composing together several functions and they can be associated with a directed acyclic graph. We may think of $f^{*}(\mathbf{x})$ as being the connection of $n$ different functions in a chain, so that:
\begin{equation}
    f^{*}(\mathbf{x})=f^{(n)}(...(f^{(2)}(f^{(1)}(\mathbf{x})))),
\end{equation}
where $f^{(1)}$,$f^{(2)}$,$\, ...$ and $f^{(n)}$ are different functions that would correspond to each layer of the network. The depth of the model corresponds to the overall length of the aforementioned chain. We call the first and last model layers the input and output layers, respectively, and the rest of the layers in between the hidden layers. Furthermore, some authors may refer to the dimensionality of the hidden layers as the width of the model. 

The edges connecting two layers are characterized by the weights, $\mathbf{w}$, and biases, $\mathbf{b}$, which are trainable parameters that are optimized during the training process. The output of the previous layer, $h_{i}$, is transformed using the affine transformation $wh_{i} + b$. Each node in the layer applies a non-linear operation to the affine transformation so that the output of layer $i + 1$ is $h_{i+1} = \sigma (wh_{i} + b)$. The non-linear function $\sigma$ is usually referred to as the activation function which must be smooth. It is worth mentioning that the output layer typically takes special activation functions, different to the ones used in the hidden layers, depending on the type of problem being solved. The model output is the predicted function that is parameterized
by the weights and biases in the network.

In the case of a FCFF, the output of a given layer $i$ is passed to layer $i+1$ and every node in the current layer is connected to every node in the previous and
subsequent layer, without any jump connections or feedback loops. On the other hand, CNNs use convolutional layers where each neuron is only connected to a few nearby neurons in the previous layer. Regarding convolutional layers, the same set of weights are shared by every neuron. This connection pattern is applied to cases where the data can be interpreted as being spatial. The convolutional layer's parameters consist of a set of learnable filters or kernels. Every filter has relatively small dimensions but extends through the full depth of the input volume. A convolution is computed by sliding the filter over the output of the previous layer. At every location, a matrix multiplication is performed and sums the result onto the feature map. The feature map is the output of the filter applied to the previous layer. In general, CNNs may also have pooling layers and fully-connected layers before the output. See~\cite{millstein2018convolutional} for an introduction to CNNs. 

For both FCFFs and CNNs, the model predictions are compared to data in a loss function, and an optimization algorithm (see \cite{ml_opt}), such as stochastic gradient descent (see \cite{sgd} and \cite{DBLP:journals/corr/Ruder16}), is used to adjust all the weights and biases, $\boldsymbol{\theta}$, of the network to minimize the loss function. The process of optimizing a neural network can be challenging since it is a high-dimensional non-convex optimization problem.

Adopting neural networks for physics-based problems entails the challenge of trying to inform the network with known physical laws, since otherwise the network will simply try to memorize the patterns in the data without paying attention to the underlying physics. One of the first attempts to embed the physical and mathematical structure of the Reynolds anisotropy tensor into a neural network was the TBNN by~\cite{ling_ml}. The TBNN guaranteed Galilean and rotational invariance of the predicted Reynolds anisotropy tensor by adding an additional tensorial layer to a FCFF network which returned the most general, local eddy viscosity model described in~\cite{pope_1975}. \cite{fang2018deep} further proposed a number of techniques such as reparameterizing a FCFF network to enforce the no-slip boundary condition as a function of the normalized distance from the wall, explicitly providing $Re_{\tau}$ to the network to incorporate friction Reynolds number information into the model, and extensions to allow for non-locality. In the present work, some of the techniques by~\cite{fang2018deep} will be applied to CNNs.

\subsection{Interpretability}
\label{subsec:Interpretability}

One of the biggest problems regarding artificial intelligence's (AI) public trust and acceptance is that most algorithms exhibit what is known as a black box behaviour. A black box algorithm lets the user see the input and output, but it gives no view of the learning process and does not explicitly share how and why the model reaches a given conclusion. Particularly in the case of DL, there may be a great number of hidden layers which allow the algorithm to automatically determine complex patterns in the input data but which obscure the inner functioning of the network and the relationship between input and output. Interpretability of neural networks aims at addressing this problem by shedding some light on the behaviour of the algorithms.

Understanding how algorithms work and make decisions is of paramount importance if we want to be able to rely on AI to make economical decisions, manage public health or, in the case of fluid mechanics, understand how our algorithm relates to the underlying physics problem that we are trying to solve. Interpretability is also a powerful tool that can be used to identify flaws in the model and improve its performance. It is currently a field in development and as mentioned by~\cite{molnar2019}, researchers have not even agreed on a proper definition for it yet. Some non-mathematical definitions have been proposed, such as those by~\cite{Miller2019ExplanationIA}: ``Interpretability is the degree to which a human can understand the cause of a decision.'' and~\cite{kim2016examples}: ``Interpretability is the degree to which a human can consistently predict the model's result.''

In this work, a number of interpretability techniques will be explored to understand the trained models and improve their architecture, explain their performance, and find relations between the model decisions and the underlying physics. Using different interpretability methods will help us validate our results.

\section{Neural Network Models}
\label{sec:Neural Network Models}

In this section, we review the FCFF neural network architectures proposed by~\cite{fang2018deep} to model the Reynolds anisotropy tensor for turbulent channel flow and introduce new CNN models. The following non-dimensional quantities are used by all models: the non-dimensional distance from the wall $y^{+}=\frac{u_{\tau}y}{\nu}$, where $u_{\tau}$ is the friction velocity and $\nu$ the kinetic viscosity; the normalized mean velocity gradient $\frac{d\overline{u}}{dy}^{*}=\frac{d\overline{u}}{dy}\frac{k}{\epsilon}$, where $k$ is the turbulent kinetic energy and $\epsilon$ is the turbulence dissipation rate; the normalized Reynolds anisotropy tensor, which is given by $\mathbf{b}=\frac{\mathbf{a}}{2k}$ and the friction Reynolds number, $Re_{\tau}$. In this work, the input to the models is the normalized mean velocity gradient of the flow $\frac{d\overline{u}}{dy}^{*}$ and the output of the model is the $u-v$ component of the normalized Reynolds anisotropy tensor $b_{uv}$.

\subsection{Review of Fully-Connected Feedforward Models for Turbulent Channel Flow}
\label{subsec:Review of MLP models}
\cite{fang2018deep} proposed a generic FCFF model as the simple baseline neural network model for their work and suggested three modifications to embed the physics of the turbulent channel flow problem into the network: to enforce the boundary condition $b_{uv}=0$ at the wall through a reparameterization, explicitly providing $Re_{\tau}$ to the network, and extensions to allow for non-local models. The present work will focus on the implementation of the first two techniques which proved to give the best results for turbulent channel flow.

\begin{figure*}
        \centering
        \begin{subfigure}[b]{0.475\textwidth}
            \centering
            \includegraphics[width=\textwidth]{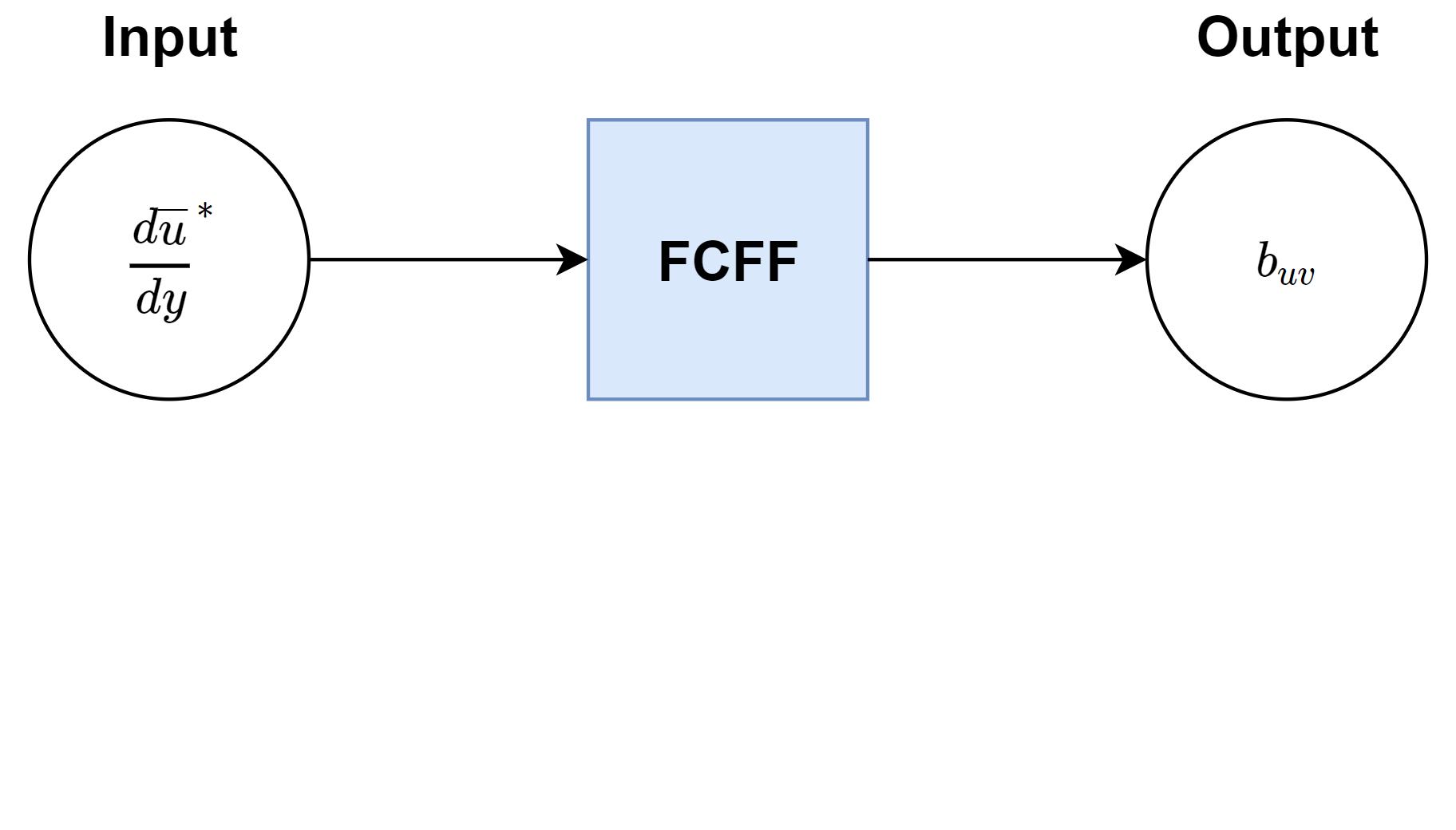}
            \caption[]%
            {{\small Baseline FCFF model.}}    
            \label{fig:FCFF_baseline}
        \end{subfigure}
        \hfill
        \begin{subfigure}[b]{0.475\textwidth}  
            \centering 
            \includegraphics[width=\textwidth]{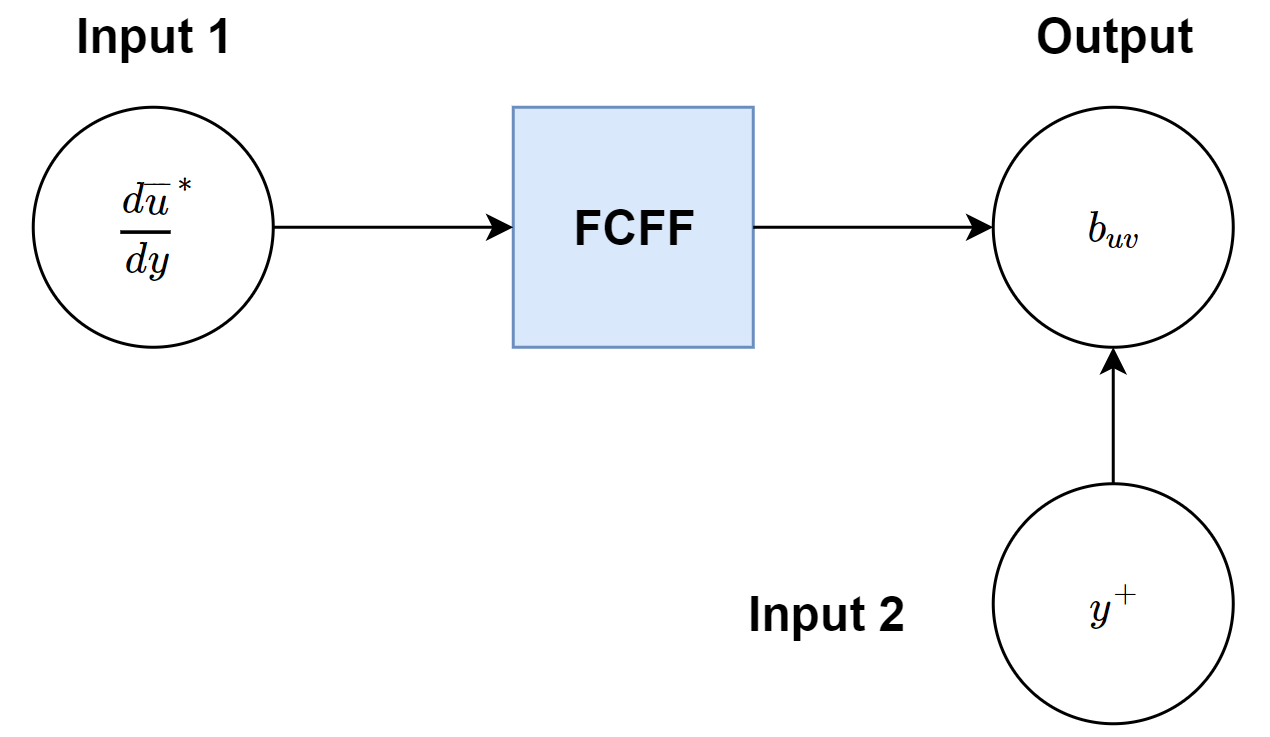}
            \caption[]%
            {{\small FCFF-BC.}}    
            \label{fig:FCFF-BC}
        \end{subfigure}
        \vskip\baselineskip
        \begin{subfigure}[b]{0.475\textwidth}   
            \centering 
            \includegraphics[width=\textwidth]{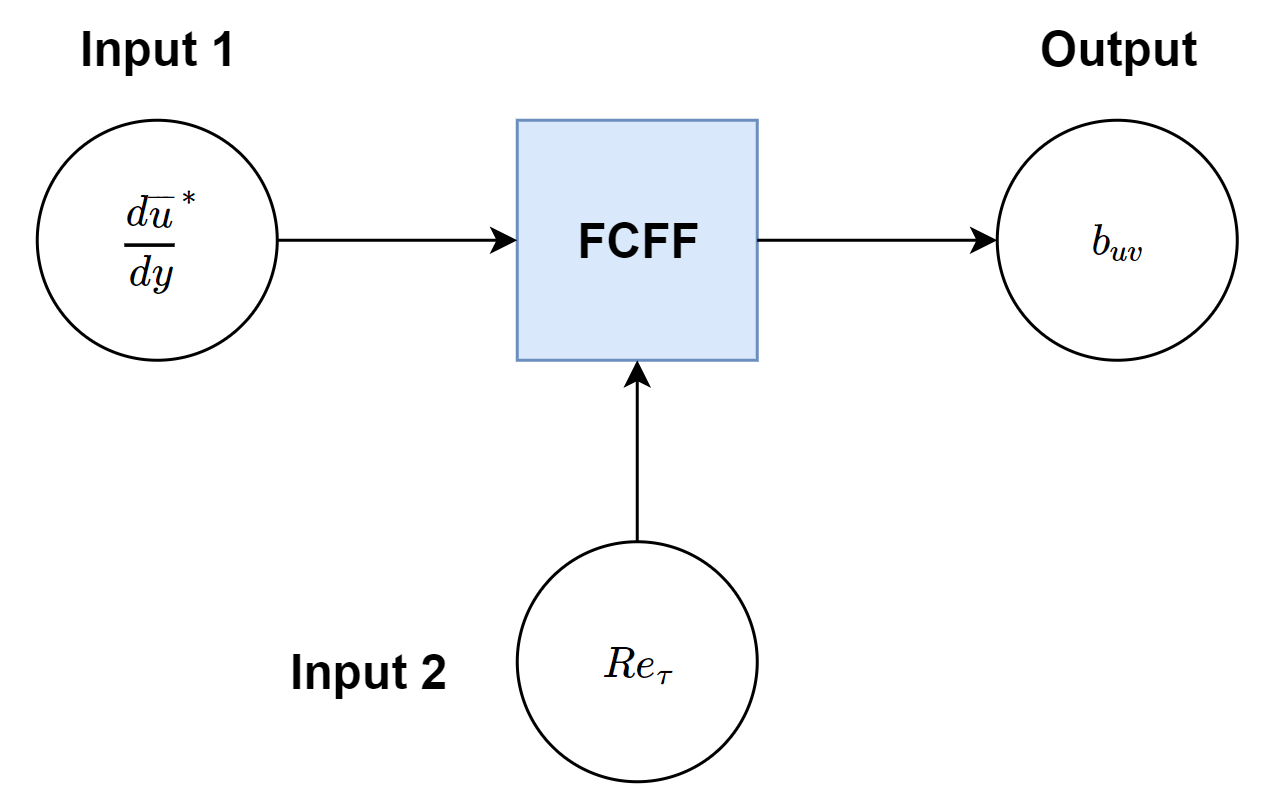}
            \caption[]%
            {{\small FCFF-$Re_{\tau}$.}}    
            \label{fig:FCFF-RE}
        \end{subfigure}
        \hfill
        \begin{subfigure}[b]{0.475\textwidth}   
            \centering 
            \includegraphics[width=\textwidth]{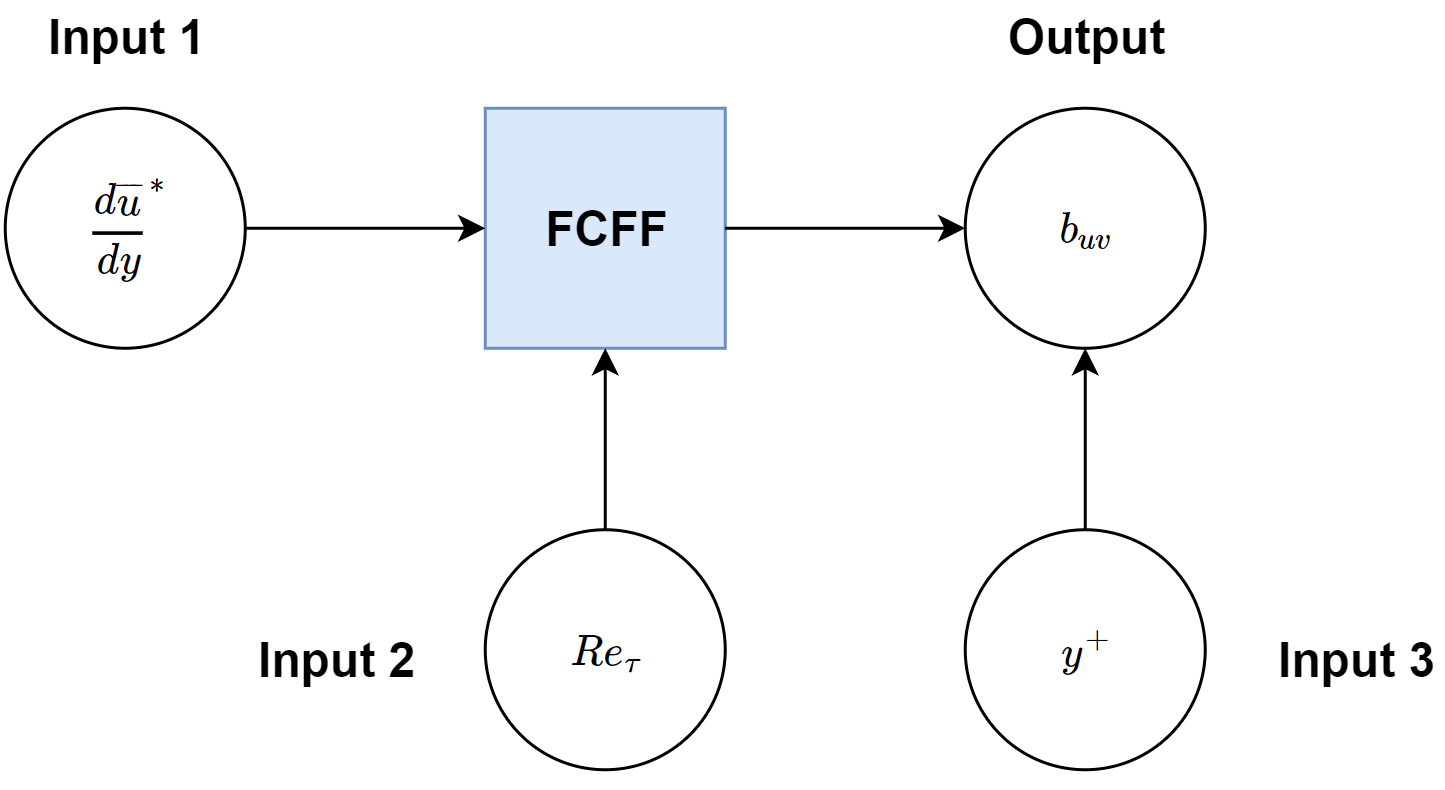}
            \caption[]%
            {{\small FCFF-BC-$Re_{\tau}$.}}    
            \label{fig:FCFF-BC-RE}
        \end{subfigure}
        \caption[]
        {\small Diagrams of FCFF models for turbulent channel flow. (a) Baseline FCFF model. (b) FCFF model with boundary condition enforcement. (c) FCFF model with friction Reynolds number injection. (d) FCFF model with boundary condition enforcement and friction Reynolds number injection.} 
        \label{fig:FCFF All plots}
    \end{figure*}

The baseline FCFF model is rather simple and has no embedded physics other than Galilean invariance since it uses the normalized mean velocity gradients as input. Figure \ref{fig:FCFF_baseline} shows a schematic representation of the baseline model, which takes the normalized mean velocity gradient at a given distance from the wall and predicts the $u-v$ component of the normalized Reynolds anisotropy tensor at that location. Note that both input and output are scalars.


The first improvement that~\cite{fang2018deep} proposed for the baseline model was enforcing the no-slip boundary condition at the channel wall by reparameterizing the output as follows,
\begin{equation}
    b_{uv}\left(\frac{d\overline{u}}{dy}^{*},y^{+}\right)=A\left(y^{+}\right)\text{FCFF}\left(\frac{d\overline{u}}{dy}^{*}\right) , 
\end{equation}
where $\text{FCFF}\left(\frac{d\overline{u}}{dy}^{*}\right)$ is the output of the baseline model and $A\left(y^{+}\right)$ is a user-selected function that must satisfy the condition $A\left(0\right)=0$. The function $A\left(y^{+}\right)=1-e^{-\beta y^{+}}$ with hyperparameter $\beta$ was suggested in the original paper, which guarantees that the solution satisfies the condition $b_{uv}=0$ at $y^{+}=0$. The improved model takes $y^{+}$ as an additional input and the loss function is calculated based on the final reparameterized solution. Figure \ref{fig:FCFF-BC} displays a diagram of the reparameterized model with the boundary condition enforcement.


\cite{fang2018deep} also suggested inputting the friction Reynolds number, $Re_{\tau}$, of the channel flow directly into the model as the friction Reynolds number has a direct influence on the mean velocity profile and, therefore, on the anisotropic Reynolds stress tensor. $Re_{\tau}$ may be fed into one or more of the intermediate layers of the FCFF model as an additional input, as shown in Figure \ref{fig:FCFF-RE}. In practice, the difference between inputting the friction Reynolds number in different layers was deemed insignificant. Furthermore, combining both boundary condition enforcement and friction Reynolds number injection as shown in Figure~\ref{fig:FCFF-BC-RE} was found to give even better results.



\subsection{Convolutional Neural Network Models}
\label{subsec: Fully Convolutional Neural Network models}
\begin{figure}[htb!]
\centering
\includegraphics[scale=0.75]{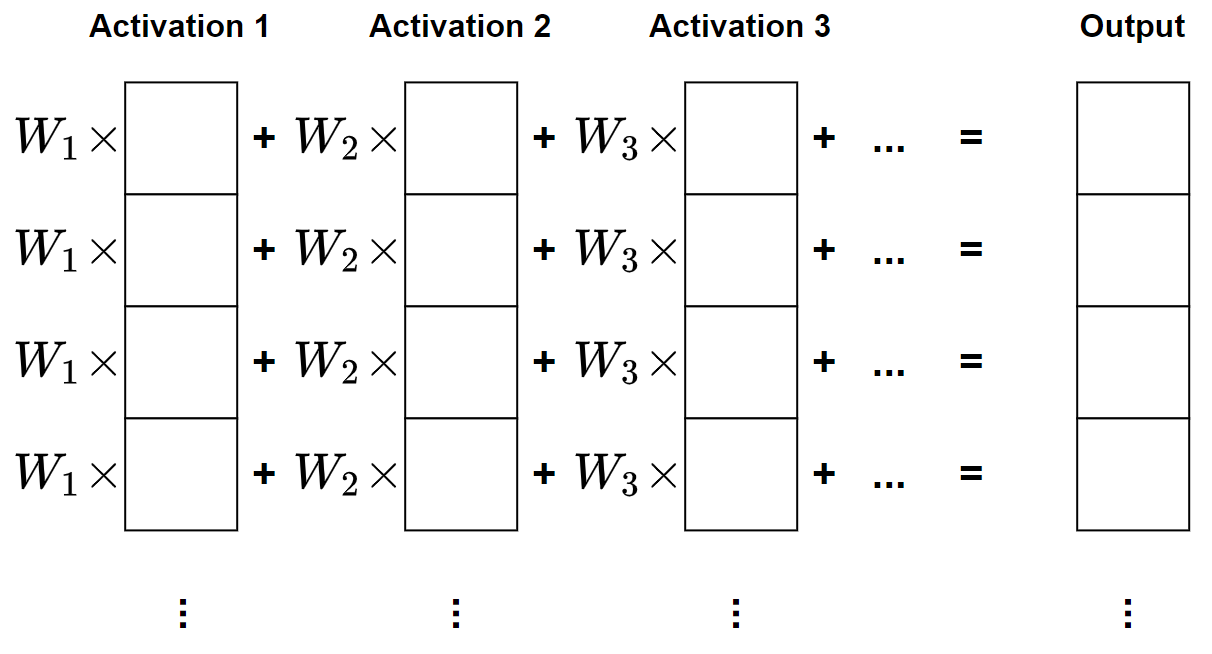}
\caption{Diagram of the final weighted sum operation. Each activation is multiplied by a trainable weight.}
\label{fig:ws}
\end{figure}

\begin{figure}[htb!]
\centering
\includegraphics[scale=0.75]{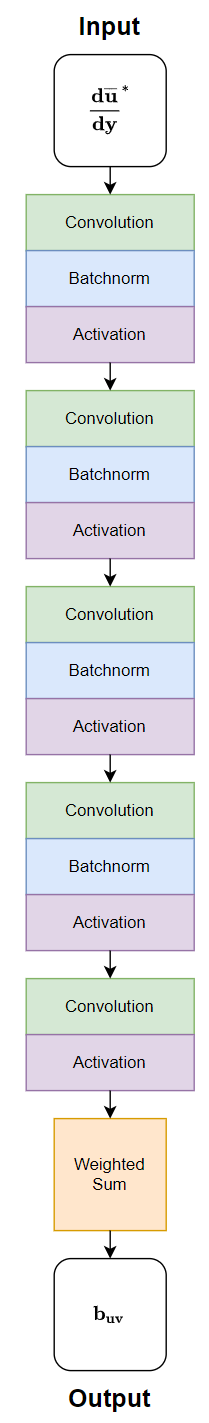}
\caption{Diagram of the inner architecture of the baseline CNN model for turbulent one-dimensional flows. The input is an array for the normalized velocity gradient profile evaluated at different locations along the height of the channel for a given friction Reynolds number. The output is an array with the corresponding predicted normalized anisotropy tensor values. The one-dimensional convolutional layers are shown in green, the one-dimensional batch normalization layers with trainable parameters are shown in blue, the activation functions in purple and the final weighted sum operation across the channel dimension in orange.}
\label{fig:CNN_baseline}
\end{figure}

In this work, we present new one-dimensional CNN models that outperform the original FCFF models previously described for turbulent channel flow. We propose a new CNN baseline model which is fully convolutional, without pooling layers, and a fully-connected region. The proposed model, which we describe in more detail in Section~\ref{subsec:Model results for Turbulent Channel Flow}, has multiple layers and multiple filters. The CNN model takes as input an array with all the normalized mean velocity gradient values along the height of the channel for a given friction Reynolds number and predicts all the normalized anisotropic Reynolds stress tensor values for that friction Reynolds number simultaneously. This is unlike the FCFF model which takes a single value of the normalized mean velocity gradient and makes a single prediction at a time. Padding is used to preserve the original length of the input array throughout all the convolutional layers, and batch normalization helps the network to train faster and be more likely to be stable (see~\cite{pmlr-v37-ioffe15}). The final convolutional layer produces several activations with the same length as the original input array. The weighted sum of the activations is calculated to obtain the final output as shown in Figure~\ref{fig:ws}. Note that the baseline CNN model also guarantees Galilean invariance since it uses the normalized mean velocity gradients as inputs. Figure~\ref{fig:CNN_baseline} shows the inner architecture of the CNN baseline model.

\begin{figure*}
        \centering
        \begin{subfigure}[b]{0.475\textwidth}
            \centering
            \includegraphics[width=\textwidth]{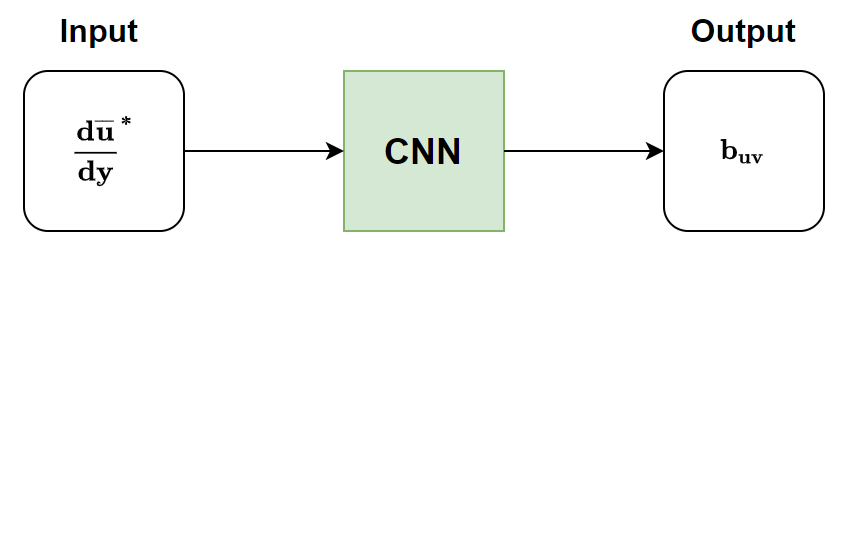}
            \caption[]%
            {{\small Baseline CNN model.}}    
            \label{fig:CNN_baseline_schematic}
        \end{subfigure}
        \hfill
        \begin{subfigure}[b]{0.475\textwidth}  
            \centering 
            \includegraphics[width=\textwidth]{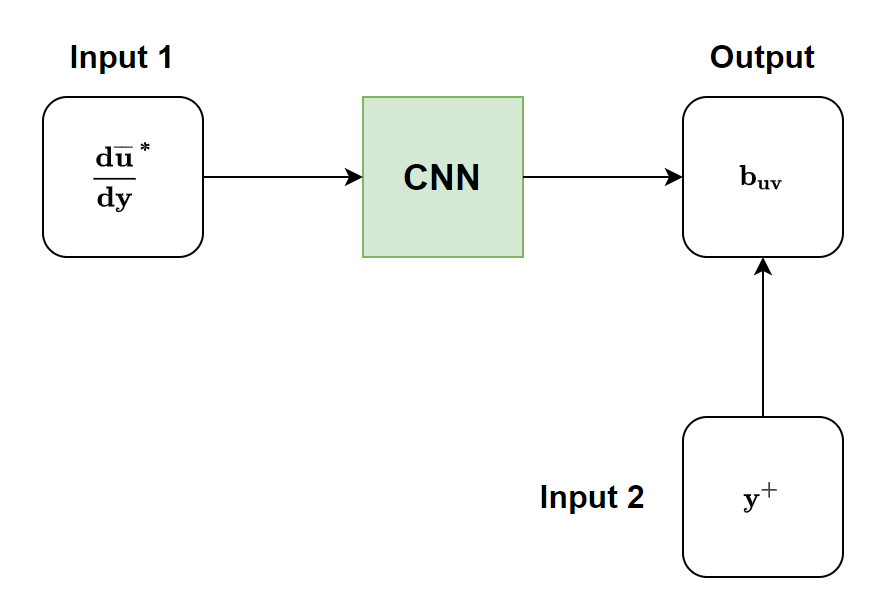}
            \caption[]%
            {{\small CNN-BC.}}    
            \label{fig:CNN-BC}
        \end{subfigure}
        \vskip\baselineskip
        \begin{subfigure}[b]{0.475\textwidth}   
            \centering 
            \includegraphics[width=\textwidth]{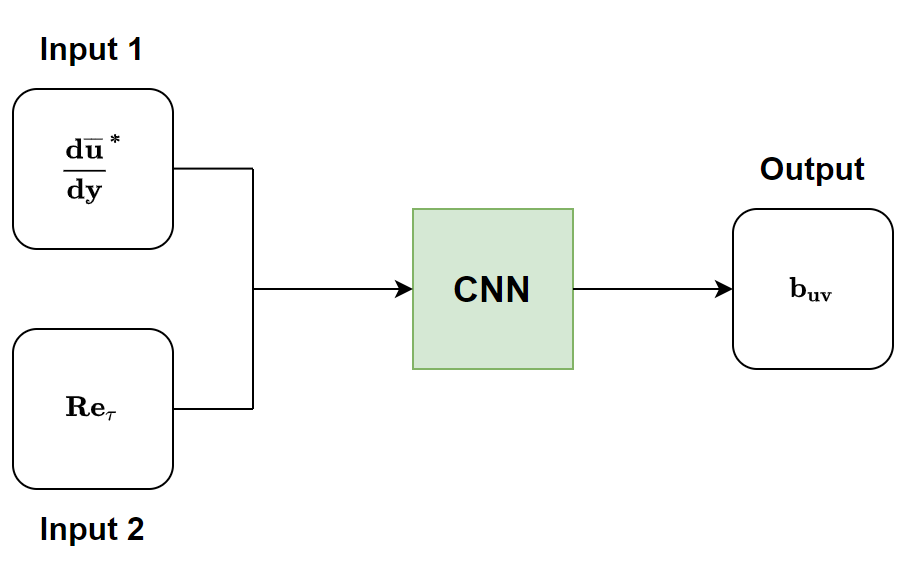}
            \caption[]%
            {{\small CNN-$Re_{\tau}$.}}    
            \label{fig:CNN-RE}
        \end{subfigure}
        \hfill
        \begin{subfigure}[b]{0.475\textwidth}   
            \centering 
            \includegraphics[width=\textwidth]{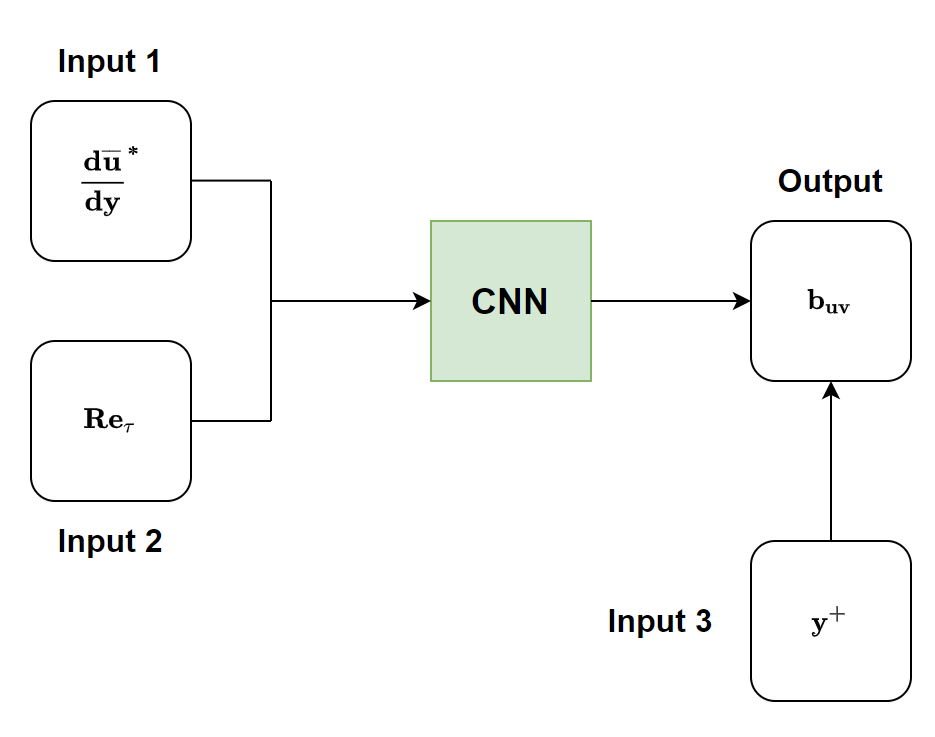}
            \caption[]%
            {{\small CNN-BC-$Re_{\tau}$.}}    
            \label{fig:CNN-BC-RE}
        \end{subfigure}
        \caption[]
        {\small Diagrams of CNN models for turbulent one-dimensional flows. (a) Baseline CNN model. (b) CNN model with boundary condition enforcement. (c) CNN model with friction Reynolds number injection. (d) CNN model with boundary condition enforcement and friction Reynolds number injection.} 
        \label{fig:CNN All plots}
    \end{figure*}

The no-slip boundary condition at the channel wall can also be enforced using the same approach as with the FCFF model,
\begin{equation}
    \mathbf{b_{uv}}\left(\mathbf{\frac{d\overline{u}}{dy}^{*},\mathbf{y^{+}}}\right)=
    A\left(\mathbf{y}^{+}\right)\text{CNN}\left(\mathbf{\frac{d\overline{u}}{dy}^{*}}\right),
\end{equation}
where $\text{CNN}\left(\mathbf{\frac{d\overline{u}}{dy}^{*}}\right)$ is the output of the baseline CNN model, and $\mathbf{y^{+}}$ and $A\left(\mathbf{y^{+}}\right)$ are the distance from the wall in viscous units and the user-selected function as described before in Section~\ref{subsec:Review of MLP models}. Figure~\ref{fig:CNN-BC} shows the diagram of the CNN model with boundary condition enforcement.


To input the friction Reynolds number into the CNN model we concatenate an array with the friction Reynolds number to the original input array with the normalized velocity gradients. The new input is a tensor with two channels, one with the non-dimensionalized velocity gradients evaluated at all the points at different locations along the height of the channel for a given friction Reynolds number, and the other with the friction Reynolds number information and all entries in the second channel being equal to the friction Reynolds number, as shown in Figure~\ref{fig:CNN-RE}. As in the case of the FCFF models, boundary condition enforcement and friction Reynolds number injection can simultaneously be applied to the CNN model (Figure~\ref{fig:CNN-BC-RE}).



\section{Interpretability Techniques}
\label{sec:Interpretability Techniques}

In the context of turbulence modelling, developing interpretability techniques will help better understand the relation between the model decisions and the underlying physics of the problem as well as to improve the model architecture. In this section, we will discuss a number of interpretability techniques that we will employ to explain and improve our CNN turbulence models.

\subsection{Activation Visualization}
\label{subsec: Activation Visualization}

Activation visualization is a straightforward technique that gives some insight into how the layer structure of a CNN model processes the data. As discussed by~\cite{deeplearningwithpython}, activation visualization consists of displaying the feature maps that are output by each layer of the model to visualize how successive convolutional layers apply transformations to the input, giving the user a high-level understanding of the meaning of each individual convolutional filter. The output of each filter may to some degree be attributed to visual or physical representations of the input data that the model is trying to capture.

In this paper, we show how to use activation visualization as a neural network design driver for our one-dimensional regression problem (see Section~\ref{subsubsec:Activation Visualization Results} for examples). By looking at the activations of the different convolutional layers within a network it is possible to determine whether some of the filters are redundant (because they are not capturing any new relationship) or more are needed (if we see the layer is struggling to learn a smooth mapping between the input and the output). In this way we can ease common issues such as underfitting and overfitting. 







\subsection{Occlusion Sensitivity}
\label{subsec: Occlusion Sensitivity}

Occlusion sensitivity was first explored by authors such as~\cite{zeiler2013visualizing} in the image classification context. In their work, the authors tried to evaluate whether the machine learning model was genuinely identifying the position of the relevant object in the image, or if it was merely paying attention to the surrounding context instead. To determine this question, they occluded different portions of the input image by sliding a grey square over it and monitored the output of the classifier. \cite{Ancona2017AUV} performed similar experiments to determine which parts of an input cause a machine learning model to change its outputs the most.

We extend this technique to our one-dimensional regression problem. We systematically occlude the input array by setting the value of the mean velocity gradients to zero at specific regions along the channel height using a sliding window and evaluate how this affects the accuracy of the prediction. We construct two different types of plots: 1.) one-dimensional global occlusion sensitivity plots that evaluate the effect of occluding different regions of the input on the overall loss of the model output and 2.) two-dimensional local occlusion sensitivity plots that aim at describing the impact of occlusion on the individual accuracy of all entries of the output array. To evaluate the loss in accuracy for the global occlusion sensitivity plots, we look at the absolute difference between the original loss of prediction and the loss of the model output when the input is partially occluded,
\begin{equation}
     \mathcal{D}_{g}=\left| L\left( \mathbf{y^{\text{true}}},\mathbf{y} \right)-L\left( \mathbf{y^{\text{true}}},\mathbf{y^{\text{occlusion}}}\right) \right|,
\label{eq:occlusion_global}
\end{equation}
where $\mathbf{y}^{\text{true}}$ is the normalized anisotropic Reynolds stress tensor across the entire domain.
For the local occlusion sensitivity plot, we individually evaluate the absolute difference of the prediction accuracy of a single output entry as follows:
\begin{equation}
    \mathcal{D}_{l}=\left| L\left( y_{i}^{\text{true}},y_{i} \right)-L\left( y_{i}^{\text{true}},y_{i}^{\text{occlusion}}\right) \right|.
\label{eq:occlusion_local}
\end{equation}
In~\eqref{eq:occlusion_global} and~\eqref{eq:occlusion_local}, $L$ is the loss function used to train the algorithm. In~\eqref{eq:occlusion_local}, $y_{i}^{\text{true}}$ corresponds to the normalized anisotropic Reynolds stress tensor value at a given position along the channel height, $y_{i}$ is the original model prediction for the anisotropic Reynolds stress tensor at that position, and $y_{i}^{\text{occlusion}}$ refers to the prediction when some of the input entries are occluded. Based on these metrics we are able to evaluate which portions of the input array are most important to preserve the accuracy of the original prediction by looking at the global loss and local loss difference when we occlude regions of the input.

\subsection{Gradient-based Sensitivity}
\label{subsec: Gradient-based Sensitivity}
\cite{simonyan2014deep} introduced the first gradient-based saliency map for image classification. This approach is often referred to as the vanilla gradient-based saliency map, which they obtained by computing the gradient of the class score with respect to the input image. For image classification, the intuition is that by calculating the change in the predicted class by applying small perturbations to the pixel values across the image, it is possible to measure the relative importance of each pixel to the final model prediction.

In our context, we calculate the absolute value of the partial derivative of the loss. The gradient provides a convenient mechanism to determine the sensitivity of the output with respect to the input. However, the main problem with gradient-based sensitivity maps in the case of image classification and regression is that they tend to be noisy.

Researchers have hypothesized about the apparent noise in raw
gradient visualization and suggested that using
the raw gradients as a measure for feature importance is not
optimal. \cite{smilkov2017smoothgrad} argued that the noise present in sensitivity maps may arise due to meaningless local variations in partial derivatives, especially given that many modern DL models use ReLU activation functions or other similar functions like ELU or Leaky ReLU. Such functions are not continuously differentiable which may lead to abrupt changes in the gradients. Furthermore,~\cite{smilkov2017smoothgrad} showed that applying a Gaussian kernel to the gradients gives more interpretable results that filter out some of the noise. 

Note that in the case of gradient-based sensitivity we can also calculate the global sensitivity by calculating the partial derivative of the overall loss with respect to the input,
\begin{equation}
    \mathcal{G}_{g}=\left|\frac{\partial\,L\left( \mathbf{y^{\text{true}}},\mathbf{y} \right)}{\partial\,\mathbf{x}}\right|,
\label{eq:grad_global}
\end{equation}
or the local sensitivity by calculating the partial derivative of the loss of an individual entry of the output with respect to the input,
\begin{equation}
    \mathcal{G}_{l}=\left|\frac{\partial\,L\left( y_{i}^{\text{true}},y_{i} \right)}{\partial\,\mathbf{x}}\right|.
\label{eq:grad_local}
\end{equation}
However, results obtained using~\eqref{eq:grad_local} tend to be too noisy and difficult to interpret, so we therefore use the global gradients in this work.

\section{Results}
\label{sec:Results}

In this section, we first discuss the model results and compare the performance of the original FCFF models by~\cite{fang2018deep} with the new CNN models. Next, we apply the interpretability techniques described in Section \ref{sec:Interpretability Techniques} to the CNN models.

\subsection{Model Results for Turbulent Channel Flow}
\label{subsec:Model results for Turbulent Channel Flow}

Table~\ref{tab:training-prediction cases} displays the four training prediction cases that are examined for turbulent channel flow. The models are trained on the data for three friction Reynolds numbers and tested on the data for the remaining friction Reynolds number. Although the same dataset is used to train both the baseline FCFF model and the baseline CNN, the training procedure is not the same for the two types of models as they are inherently different. The FCFF model by~\cite{fang2018deep} maps a single scalar input to a scalar output,
\begin{equation}
    b_{uv}=\text{FCFF}\left(\frac{d\overline{u}}{dy}^{*}\right)\,,
\end{equation}
whereas the CNN model maps an input array to an output array,
\begin{equation}
    \mathbf{b_{uv}}= \text{CNN}\left(\mathbf{\frac{d\overline{u}}{dy}^{*}}\right).
\end{equation}
The FCFF model implicitly assumes that the normalized Reynolds stress tensor is only dependent on the local normalized mean velocity gradient, which we know to be false. \cite{fang2018deep} tried to address this problem by introducing non-locality using fractional derivatives but it was not as satisfactory as expected. On the other hand, because the CNN model proposed in the present work takes the whole normalized mean velocity gradient as input, it is more effective at capturing non-local relationships.

\begin{table}[htb!]
\centering
\caption{The four training-prediction cases for turbulent channel flow.}
\label{tab:training-prediction cases}
\begin{tabular}{@{}ccc@{}}
\toprule
Case & Training set            & Test set \\ \midrule
1    & $Re_{\tau}={[}550,1000,2000{]}$  & $Re_{\tau}=5200$  \\
2    & $Re_{\tau}={[}550,1000,5200{]}$  & $Re_{\tau}=2000$  \\
3    & $Re_{\tau}={[}550,2000,5200{]}$  & $Re_{\tau}=1000$  \\
4    & $Re_{\tau}={[}1000,2000,5200{]}$ & $Re_{\tau}=550$ \\ \bottomrule 
\end{tabular}
\end{table}

\begin{figure}[htb!]
\centering
\includegraphics[width=0.65\linewidth]{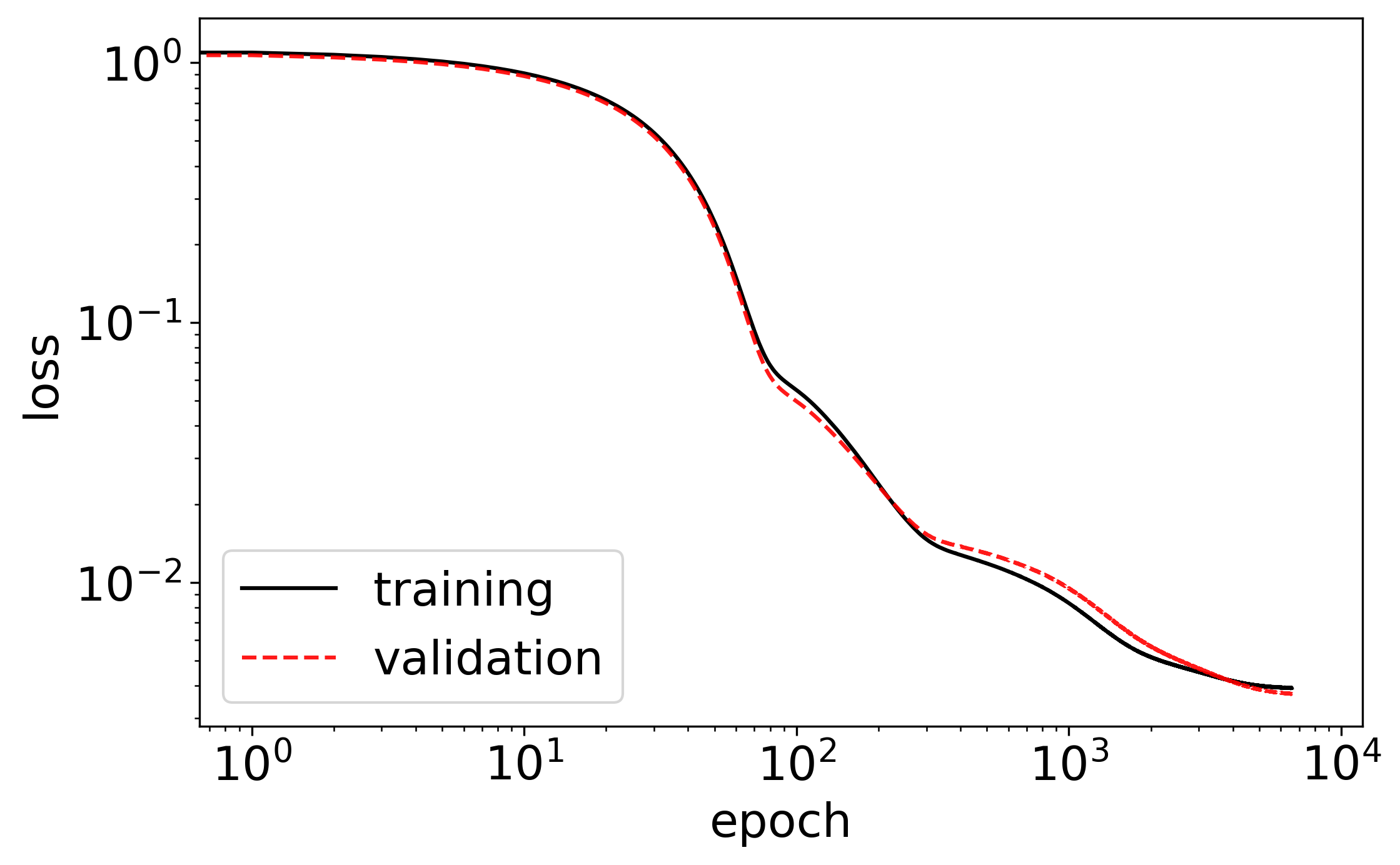}
\caption{The training and validation loss as a function of the number of epochs during training for the FCFF-BC-$Re_{\tau}$ model
for Case 2. The model converged after 6577 epochs.}
\label{fig:loss-MLP-BC-RE}
\end{figure}

All FCFF models had 5 layers with 50 nodes each and ELU activation functions as in the paper by~\cite{fang2018deep}. To train the FCFF models we randomly split and shuffled the training
data into $80\%$ training and $20\%$ validation data. We empirically found that training the model to predict ten times the normalized anisotropic Reynolds stress tensor, $10b_{uv}$, helped with model convergence, and we simply divide the model output of the trained model by $10$ to make predictions. The mean squared error between the labels and
the predicted value of the normalized anisotropic Reynolds stress tensor was used as the loss function. The training optimization was performed using the Adam optimizer~\cite{kingma2014adam} with an initial learning rate of $10^{-4}$ and a batch size of 10. The weights and biases of the networks were initialized using the Kaiming He uniform distribution as described in~\cite{he2015delving}. To prevent over-fitting we used early stopping by monitoring the validation loss as suggested by~\cite{prechelt1998early}. For the boundary condition enforcement function $A\left(y^{+}\right)=1-e^{-\beta y^{+}}$ we found $\beta=0.1$ to work best. For friction Reynolds number injection, multiplying $Re_{\tau}$ times $10^{-8}$ helped the training. The relative order of magnitude of the input is of paramount importance when training a DL algorithm for fast convergence. We found that this scaling helped substantially speed up the optimization, allowing the neural network to capture information about the friction Reynolds number while still focusing most of its attention on the velocity gradients. Figure \ref{fig:loss-MLP-BC-RE} shows the training and validation loss during training of the FCFF-BC-$Re_{\tau}$ model (FCFF model with boundary condition enforcement and friction Reynolds number injection) for Case 2 as a function of the number of epochs. In this case, training took 6577 epochs. For the rest of the FCFF models the number of epochs was of the same order of magnitude.

On the other hand, the training procedure for the CNN models was modified with respect to the FCFF models. We considered the training data as three ``images'' of the normalized mean velocity gradient profile, one for each of the friction Reynolds numbers. The three images were repeatedly fed into the model. The batch size was 1 since we fed one image to the model at a time. Given that the number of data points for different friction Reynolds numbers varies in the DNS dataset ---recall from Section \ref{subsec:Turbulent channel flow} that $Ny= [192,256,384,768]$ for $Re_{\tau}= [550,1000,2000,5200]$--- we used padding on the profiles with less data points to equalize the length of all inputs. Again, we found that training the model to predict ten times the normalized anisotropic Reynolds stress tensor, $10b_{uv}$, helped with model convergence. The mean squared error between the labels and
the predicted value of the normalized anisotropic Reynolds stress tensor was used as the loss function for the CNN models as well. However, note that instead of evaluating the loss on a single prediction, the mean squared error of the whole $\mathbf{b_{uv}}$ profile prediction for a given friction Reynolds number was evaluated all together. We also used the Adam optimizer with initial learning rate of $10^{-4}$ and the Kaiming He initialization for the convolutional layers. The CNN model had 5 convolutional layers and the first four included batch normalization with trainable parameters before applying the ELU activation function as shown in Figure~\ref{fig:CNN_baseline}. The first convolutional layer had 5 filters of kernel size 3, the second layer had 5 filter of kernel size 11, the third layer had 10 filters of kernel size 31 and the last two convolutional layers had 10 filters of kernel size 41. Given that the last convolutional layer had 10 filters, 10 weights were used in the final weighted sum. The trainable parameters for the weighted sum at the end of the model architecture were initialized at $-\frac{1}{10}$. The negative sign in the weights improved training and eased the mapping from input to output. We empirically found that using big kernel sizes in the last layers guaranteed a smooth output curve. Additionally, weight decay with hyperparameter $\lambda=10^{-4}$ (found through hyperparameter optimization) was used and observed to help obtain a smooth solution without significantly slowing down training. The boundary condition enforcement hyperparameter $\beta$ remained unaltered from the FCFF model. Once again, multiplying the friction Reynolds number times $10^{-8}$ before injection also helped the training of this model. In the case of the CNN models, training lasted 10000 to 25000 epochs for turbulent channel flow, significantly longer than in the case of the FCFF model. Alternatively, one may choose to use a deeper neural network with smaller kernel sizes. We found that to obtain similar results using a maximum kernel size of 11 the neural network had to be at least twice as deep which slowed down training. Using approximately the same number of trainable parameters in a deeper network with smaller kernel sizes implied that the algorithm had to calculate longer chain rules to update the trainable parameters even as the model was more likely to suffer from vanishing gradient problems.

For model evaluation, we used the $R^{2}$ score, which is a statistical measure that represents the proportion of the variance for a dependent variable that is explained by an independent variable in a regression model. An $R^{2}$ of $1$ indicates a perfect fit. The $R^{2}$ can be negative on the validation set if the regression does worse than the sample mean in terms of tracking the dependent variable. The $R^{2}$ scores of the various models on the training and test data are displayed in Table~\ref{tab:model_performance_r2}. 

\begin{figure}[htbp!]
\centering
\subfloat[]{\label{fig:model_results_channel_flow_1}\includegraphics[width=0.85\linewidth]{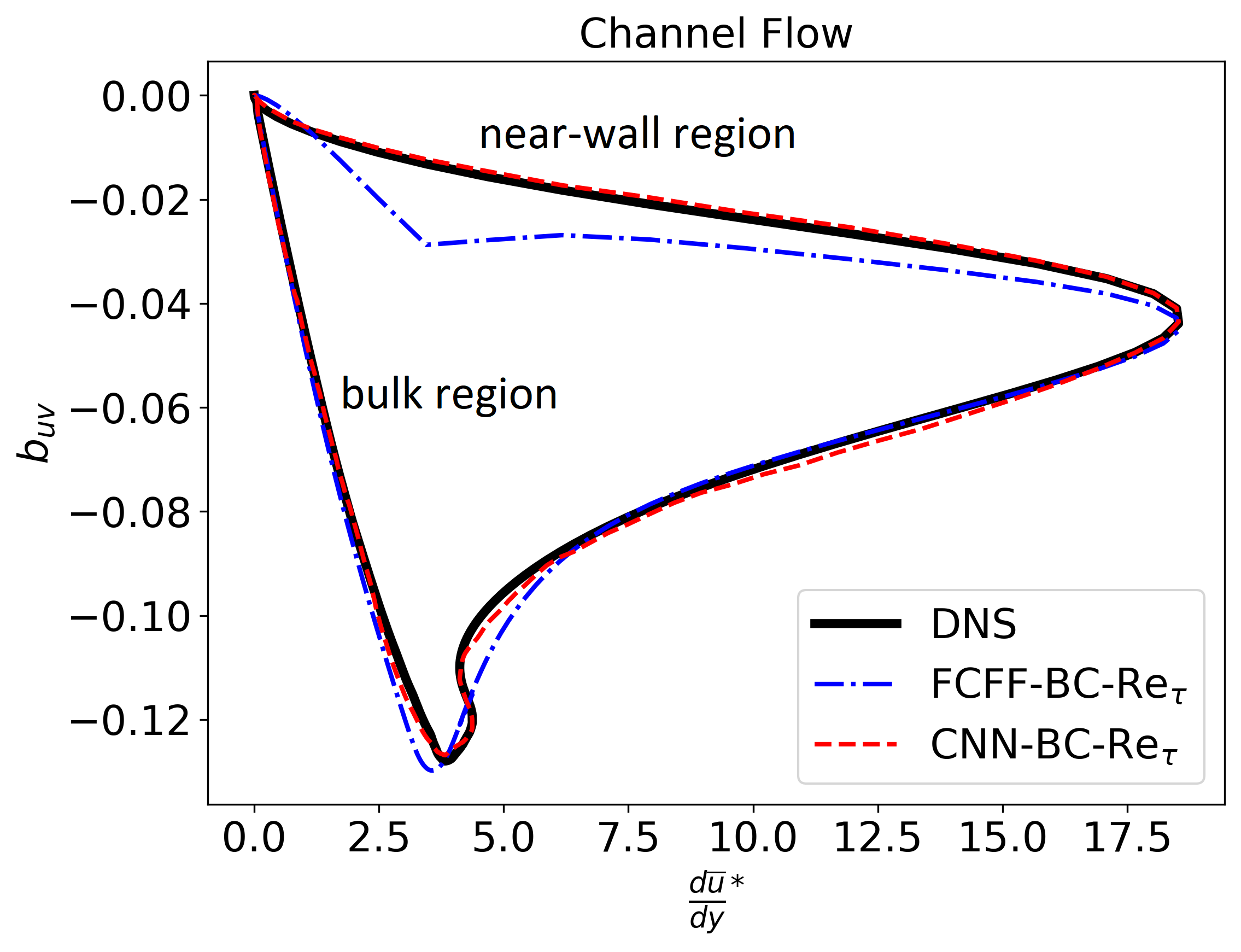}}\qquad
\subfloat[]{\label{fig:model_results_channel_flow_2}\includegraphics[width=0.85\linewidth]{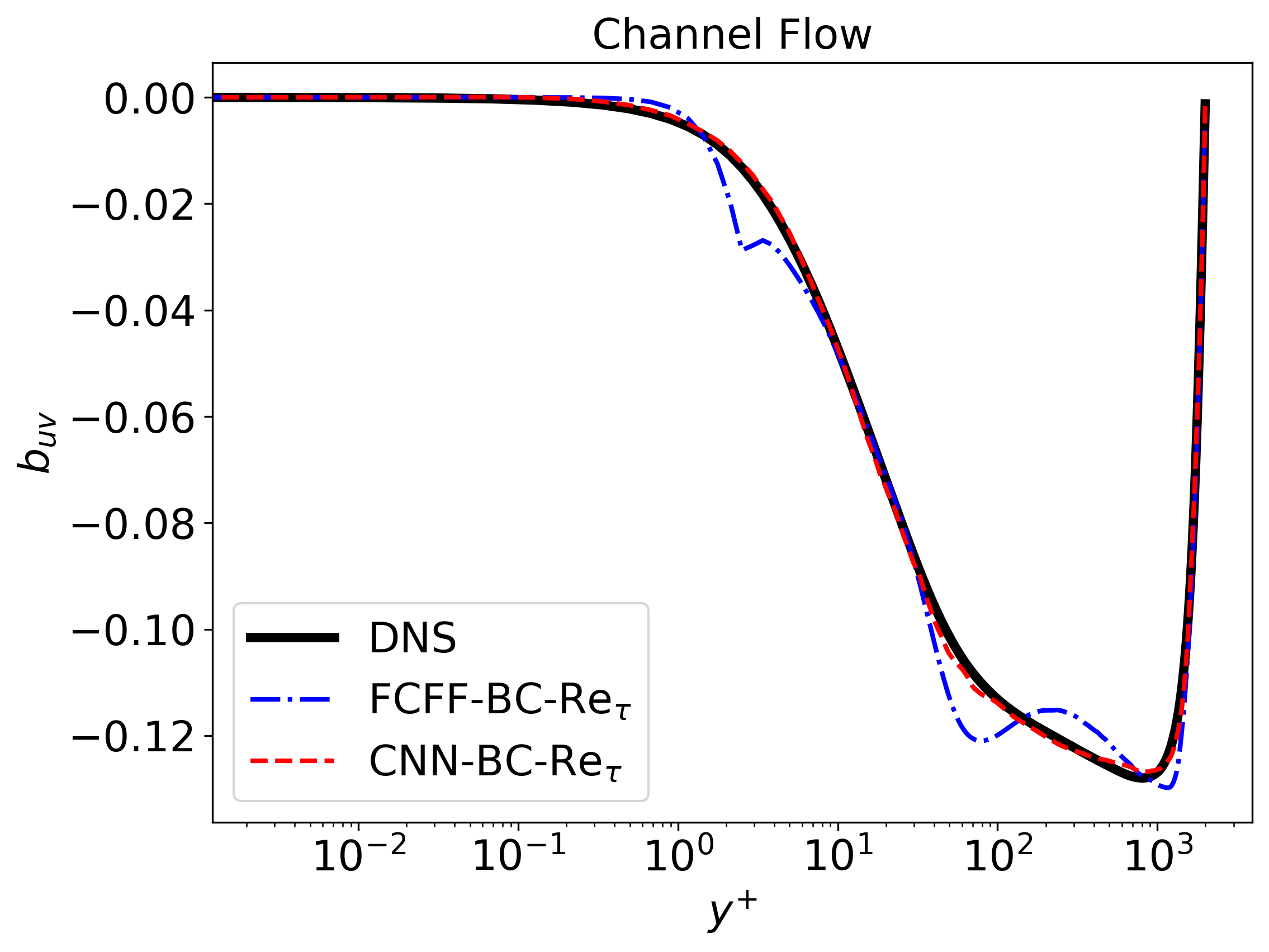}}
\caption{Comparison between FCFF-BC-$Re_\tau$ and CNN-BC-$Re_\tau$ model for turbulent channel flow Case 2 test. (a) Displays the two model predictions against the normalized mean velocity gradient profile. (b) Shows the predictions against the normalized distance from the wall in viscous units.}
\label{fig:Model_results_channel_flow}
\end{figure}

The CNN models consistently outperform the FCFF models in all cases. Figure~\ref{fig:Model_results_channel_flow} shows a comparison between the best performing FCFF and the best CNN model for turbulent channel flow Case 2. The CNN model provides an almost perfect fit to the DNS data, whereas the original FCFF model fails at some key regions. We find that both the boundary condition enforcement and the friction Reynolds number injection proposed by~\cite{fang2018deep} also prove successful in the case of CNN models and help the algorithm better learn the relationship between the normalized mean velocity gradient and the $u-v$ component of the normalized anisotropic Reynolds stress tensor. With a few exceptions, combining these two physics embedding techniques with the new baseline CNN model seems to be the best choice. Note that the best CNN model is able to outperform the best FCFF model using less parameters as shown in Table \ref{tab:number of parameters}, which confirms that the new architecture is more efficient. 

\begin{table}[htb!]
    \centering
        \caption{$R^2$ score of $b_{uv}$ predictions by various models for four turbulent channel flow training-prediction cases.}
        \begin{tabular}{@{}lllllllll@{}}
            \toprule
             & \multicolumn{2}{c}{Case 1} & \multicolumn{2}{c}{Case 2} & \multicolumn{2}{c}{Case 3} & \multicolumn{2}{c}{Case 4} \\
             \cmidrule(l){2-3} \cmidrule(l){4-5} \cmidrule(l){6-7} \cmidrule(l){8-9} 
             Model Type & Train & Test & Train & Test & Train &  Test & Train & Test \\
            \midrule
            CNN                & 0.9999 & 0.9299 & 0.9999 & 0.9949 & 0.9999 & 0.9978 & 0.9999 & 0.9894 \\
            CNN-BC & 0.9999 & 0.9069 & 0.9999 & 0.9968 & 0.9998 & 0.9991 & 0.9998 & 0.9954 \\
            CNN-$Re_\tau$     & 0.9999 & 0.9327 & 0.9997 & 0.9960 & 0.9996 & 0.9925 & 0.9998 & 0.9968 \\
            CNN-BC-$Re_\tau$      & \textbf{0.9999} & \textbf{0.9901} & \textbf{0.9999} & \textbf{0.9982} & \textbf{0.9999} & \textbf{0.9991} & \textbf{0.9999} & \textbf{0.9953} \\
            FCFF                & 0.8616 & 0.9118 & 0.8688 & 0.8853 & 0.8838 & 0.8726 & 0.8981 & 0.8004 \\
            FCFF-BC & 0.9614 & 0.9045 & 0.9645 & 0.9742 & 0.9689 & 0.9798 & 0.9825 & 0.9148 \\
            FCFF-$Re_\tau$     & 0.8633 & 0.9041 & 0.8644 & 0.8839 & 0.8788 & 0.8693 & 0.8971 & 0.8011 \\
            FCFF-BC-$Re_\tau$      & 0.9737 & 0.9473 & 0.9695 & 0.9782 & 0.9691 & 0.9814 & 0.9831 & 0.9106 \\
            \bottomrule
        \end{tabular}
        \label{tab:model_performance_r2}
    \end{table}

\begin{table}[htb!]
\centering
\caption{Number of trainable parameters for each model.}
\label{tab:number of parameters}
\begin{tabular}{@{}lc@{}}
\toprule
 Model Type & Number of Trainable Parameters\\ \midrule
CNN    & 10151\\
CNN-BC  &  10151\\
CNN-$Re_\tau$    & 10166\\
CNN-BC-$Re_\tau$    & 10166\\
FCFF    & 5351\\
FCFF-BC    & 5351\\
FCFF-$Re_\tau$    & 10401\\
FCFF-BC-$Re_\tau$    & 10401\\\bottomrule 
\end{tabular}
\end{table}

Next, we proceed to test the applicability of the CNN model with boundary condition enforcement and Reynolds number injection to other one-dimensional turbulent flows. Later, in Section \ref{subsec:Interpretability results}, we will give guidelines for the reader to use interpretability to fine-tune the model to their particular one-dimensional turbulent flow of interest.  
\subsection{Model Results for Turbulent Couette Flow}
\label{subsec:Model results for Turbulent Couette Flow}

\begin{figure}[htbp!]
\centering
\subfloat[]{\label{fig:model_results_couette_flow_1}\includegraphics[width=0.85\linewidth]{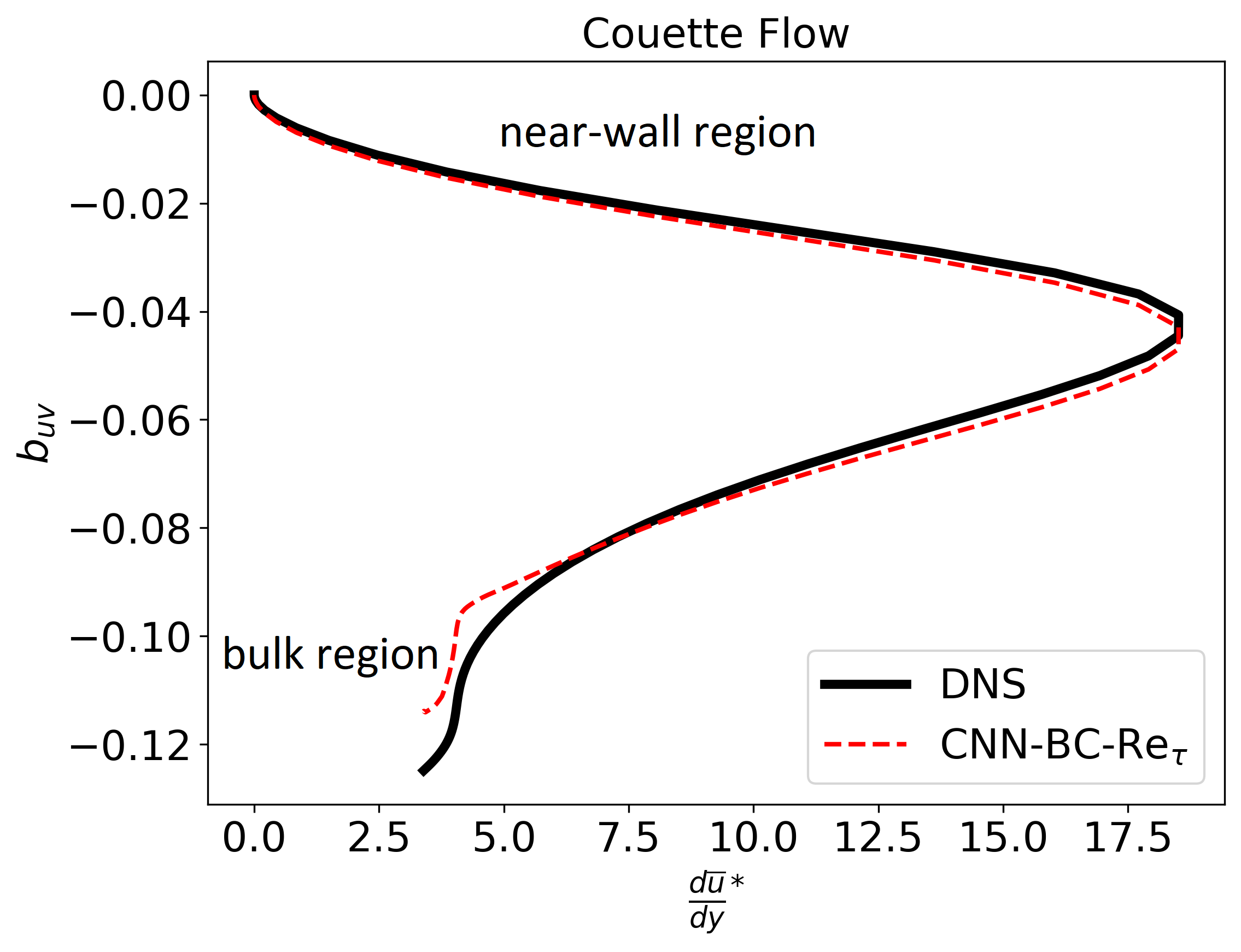}}\qquad
\subfloat[]{\label{fig:model_results_couette_flow_2}\includegraphics[width=0.85\linewidth]{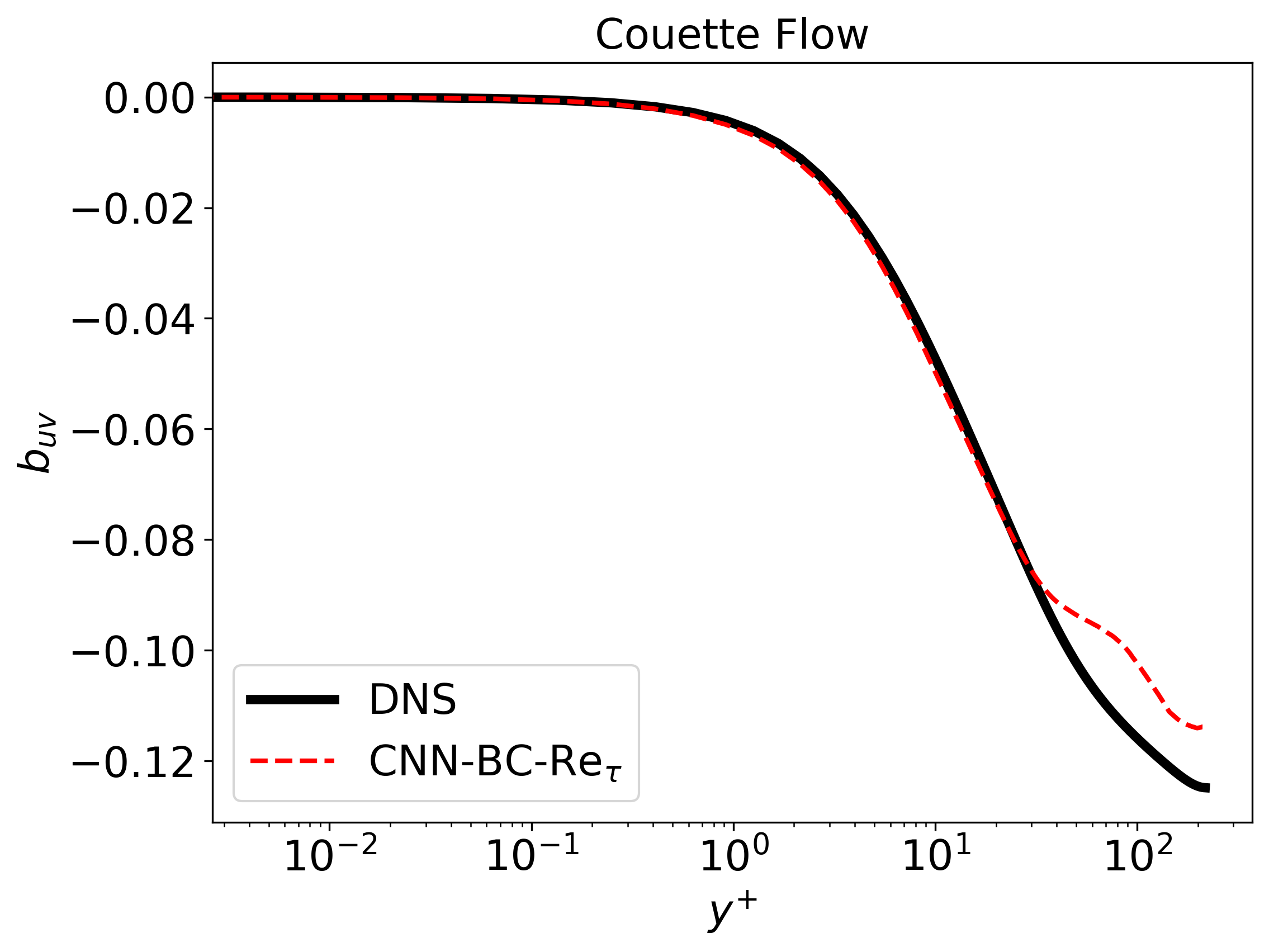}}
\caption{CNN-BC-$Re_\tau$ model prediction for turbulent Couette flow Case 2 test. (a) Displays the model predictions against the normalized mean velocity gradient profile and the DNS data. (b) Shows the prediction against the normalized distance from the wall in viscous units and the DNS data.}
\label{fig:Model_results_couette_flow}
\end{figure}

Table \ref{tab:training-prediction cases couette} displays the three training-prediction cases that are examined for turbulent Couette flow. As previously mentioned in Section \ref{subsec:Turbulent Couette flow}, the DNS data used for this flow case consists of fewer data points, for $Re_{\tau} = [93, 220, 500]$ the number of points is $N_{y}=[64,96,128]$, which makes it harder to learn the relationship. Although the model performance is slightly worse than that for turbulent channel flow, the $R^{2}$ scores shown in in Table \ref{tab:model_performance_r2_couette} are still satisfactory. Note that we are training and testing the CNN model architecture that was optimized for turbulent channel flow on a different flow configuration. As later discussed in Section~\ref{subsubsec:Activation Visualization Results}, fine-tuning the model for this specific case would help yield better results. Figure \ref{fig:Model_results_couette_flow} shows the prediction for Case 2, plotted against the normalized mean velocity gradient and against the normalized distance from the wall. Overall, we find that the predictions for Couette flow are particularly good for the near-wall region, as expected thanks to the boundary condition enforcement, but the model struggles to capture the right shape as it gets closer to the bulk.

\begin{table}[htb!]
\centering
\caption{The three training-prediction cases for turbulent Couette flow.}
\label{tab:training-prediction cases couette}
\begin{tabular}{@{}ccc@{}}
\toprule
Case & Training set            & Test set \\ \midrule
1    & $Re_{\tau}$={[}93,220{]}  & $Re_{\tau}$=500  \\
2    & $Re_{\tau}$={[}93,500{]}  & $Re_{\tau}$=220  \\
3    & $Re_{\tau}$={[}220,500{]} & $Re_{\tau}$=93 \\ \bottomrule 
\end{tabular}
\end{table}

\begin{table}[htp!]
\centering
        \caption{$R^2$ score of $b_{uv}$ predictions of the best CNN model for three turbulent Couette flow training-prediction cases.}
        \begin{tabular}{@{}lllllll@{}}
            \toprule
             & \multicolumn{2}{c}{Case 1} & \multicolumn{2}{c}{Case 2} & \multicolumn{2}{c}{Case 3} \\
             \cmidrule(l){2-3} \cmidrule(l){4-5} \cmidrule(l){6-7}  
             & Train & Test & Train & Test & Train &  Test\\
            \midrule
            CNN-BC-$Re_\tau$      & 0.9999 & 0.9223 & 0.9999 & 0.9575 & 0.9999 & 0.9708\\
            \bottomrule
        \end{tabular}
        \label{tab:model_performance_r2_couette}
    \end{table}
    
\subsection{Model Results for Turbulent Channel Flow with Wall Transpiration}
\label{subsec:Model results for Turbulent Channel Flow with Wall Transpiration}

\begin{figure}[htbp!]
\centering
\subfloat[]{\label{fig:model_results_transpiration_flow_1}\includegraphics[width=0.85\linewidth]{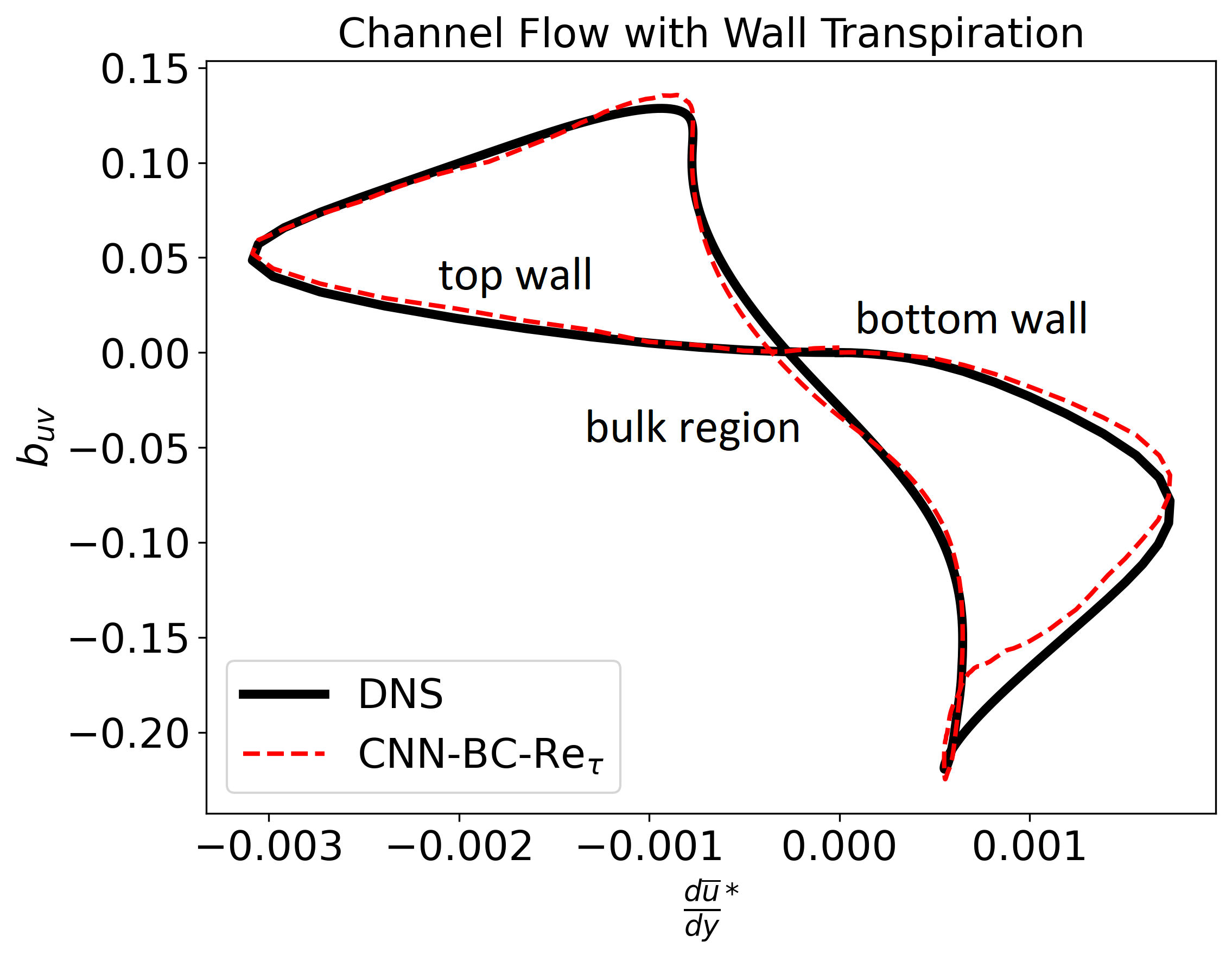}}\qquad
\subfloat[]{\label{fig:model_results_transpiration_flow_2}\includegraphics[width=0.85\linewidth]{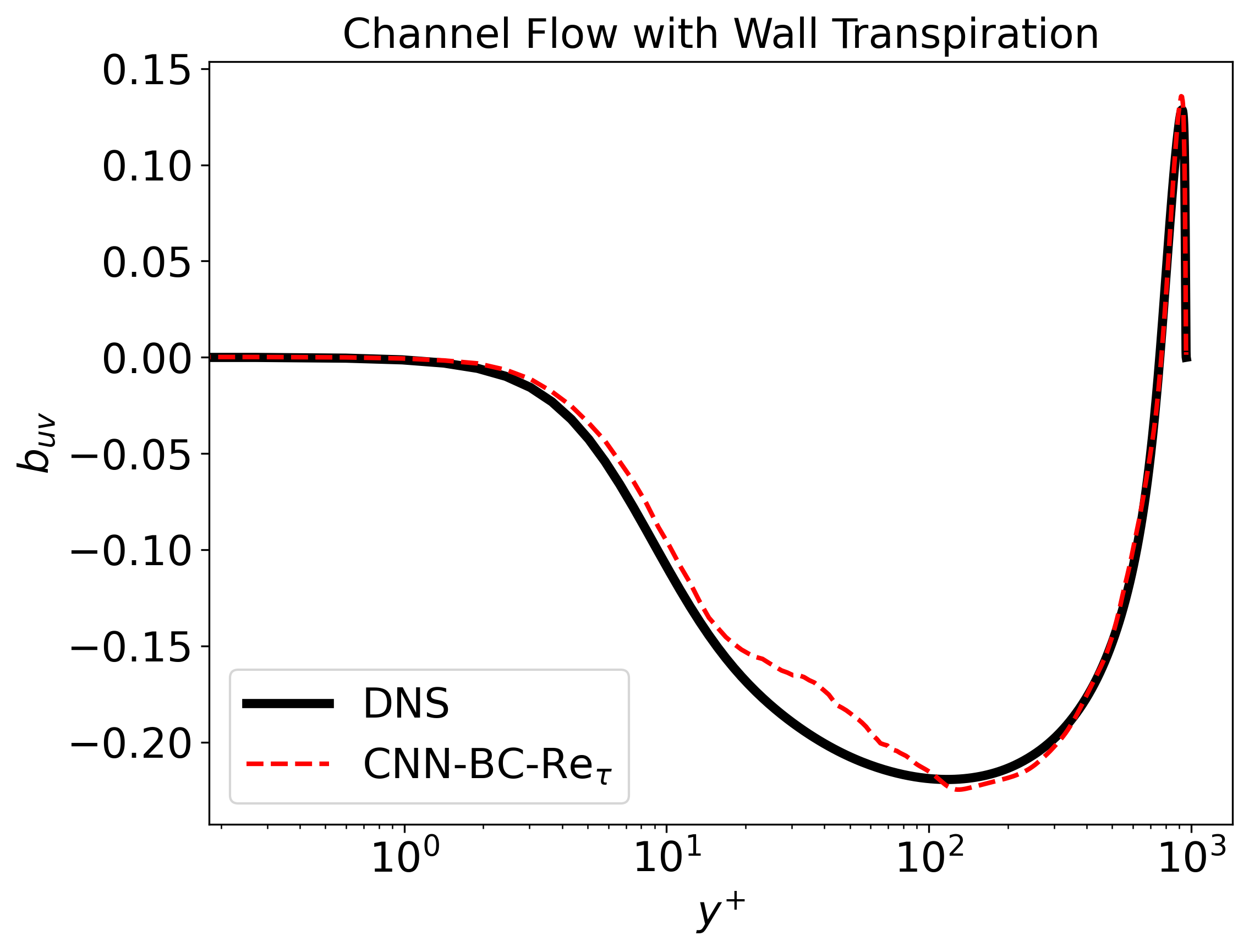}}
\caption{CNN-BC-$Re_\tau$ model prediction for turbulent channel flow with wall transpiration Case 2 test. (a) Displays the model predictions against the normalized mean velocity gradient profile and the DNS data. (b) Shows the prediction against the normalized distance from the wall in viscous units and the DNS data.}
\label{fig:Model_results_transpiration_flow}
\end{figure}

As mentioned in Section \ref{subsec:Turbulent Channel Flow with Wall Transpiration}, the DNS data used for turbulent channel flow with wall transpiration had three friction Reynolds numbers $Re_{\tau} = [250, 480, 850]$ with $N_{y}=[251,385,471]$. Table \ref{tab:training-prediction cases wall transpiration} shows the cases for this flow configuration and Table \ref{tab:model_performance_r2_transpiration} the results. Figure \ref{fig:Model_results_transpiration_flow} shows the prediction for Case 2. The introduction of a transpiration velocity substantially complicates the input-output relationship compared to the original channel flow studied by~\cite{fang2018deep}, which forces the model to learn a more complex mapping. Nevertheless, we find that our new CNN model is able to be successfully applied to this flow and capture the features of the DNS data.  

\begin{table}[htb!]
\centering
\caption{The three training-prediction cases for turbulent channel flow with wall transpiration.}
\label{tab:training-prediction cases wall transpiration}
\begin{tabular}{@{}ccc@{}}
\toprule
Case & Training set            & Test set \\ \midrule
1    & $Re_{\tau}$={[}250,480{]}  & $Re_{\tau}$=850  \\
2    & $Re_{\tau}$={[}250,850{]}  & $Re_{\tau}$=480  \\
3    & $Re_{\tau}$={[}480,850{]} & $Re_{\tau}$=250 \\ \bottomrule 
\end{tabular}
\end{table}

\begin{table}[htp]
\centering
        \caption{$R^2$ score of $b_{uv}$ predictions of the best CNN model for three turbulent channel flow with wall transpiration training-prediction cases.}
        \begin{tabular}{@{}lllllll@{}}
            \toprule
             & \multicolumn{2}{c}{Case 1} & \multicolumn{2}{c}{Case 2} & \multicolumn{2}{c}{Case 3} \\
             \cmidrule(l){2-3} \cmidrule(l){4-5} \cmidrule(l){6-7}  
             & Train & Test & Train & Test & Train &  Test\\
            \midrule
            CNN-BC-$Re_\tau$      & 0.9999 & 0.9865 & 0.9999 & 0.9949 & 0.9999 & 0.9459\\
            \bottomrule
        \end{tabular}
        \label{tab:model_performance_r2_transpiration}
    \end{table}
    
\subsection{Interpretability Results}
\label{subsec:Interpretability results}

The CNN models have shown excellent performance across a range of friction Reynolds numbers as well as their applicability to different one-dimensional turbulent flows. Now, we proceed to apply the interpretability techniques described in Section \ref{sec:Interpretability Techniques} to the models to gain a better understanding of their decision-making process and the influence of physics embedding techniques such as boundary condition enforcement and friction Reynolds number injection.

\subsubsection{Activation Visualization Results}
\label{subsubsec:Activation Visualization Results}

Activation visualization allows us to get an overall understanding of the transformations that the model applies to the input as well as when some filters have become redundant or more are needed to capture a complex behaviour.

\begin{figure}[htbp!]
\centering
\subfloat[First layer activation output.]{\label{fig:layer1}\includegraphics[width=0.475\linewidth]{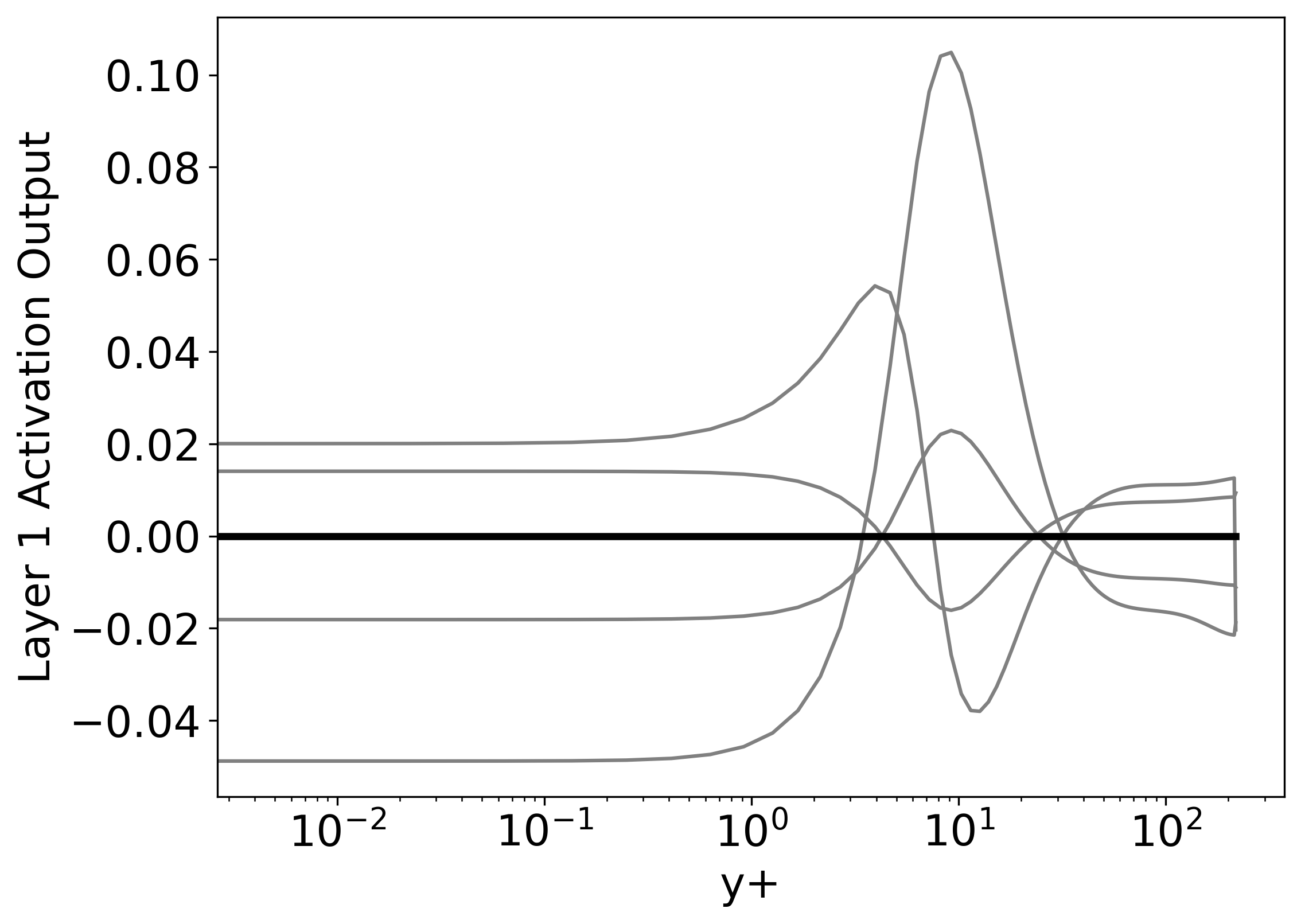}}\qquad
\subfloat[Second layer activation output.]{\label{fig:layer2}\includegraphics[width=0.475\linewidth]{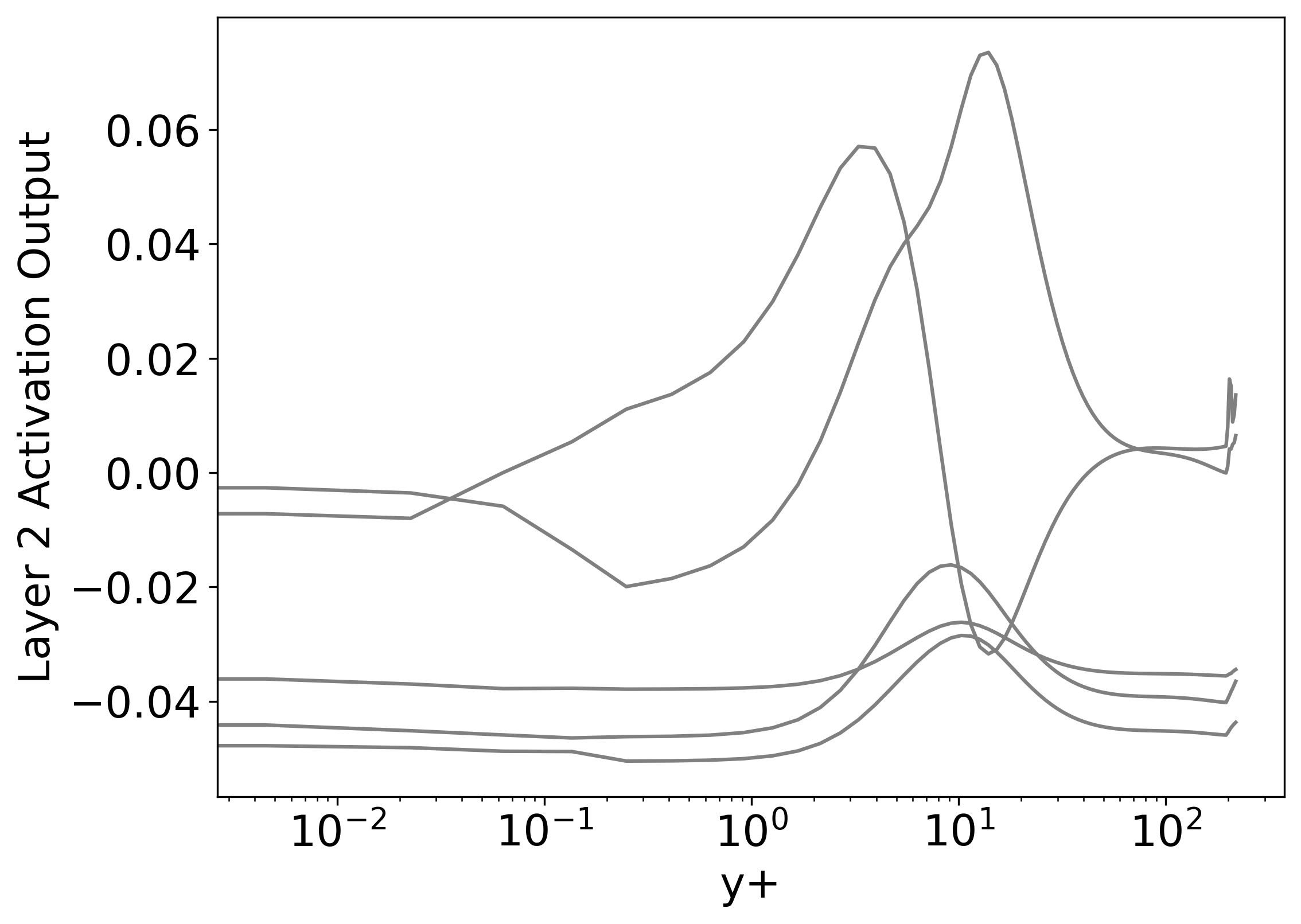}}\\
\subfloat[Third layer activation output.]{\label{fig:layer3}\includegraphics[width=0.475\textwidth]{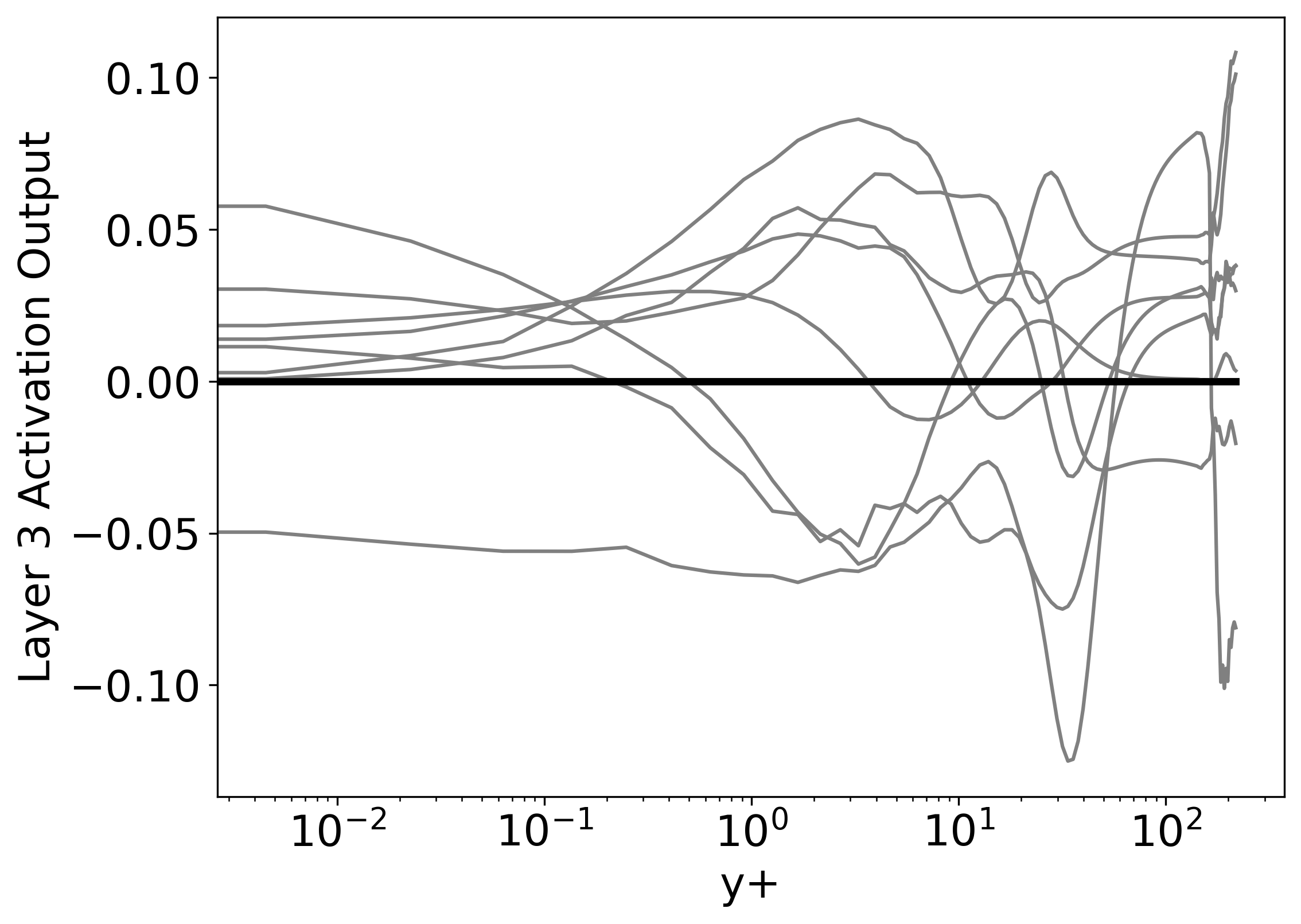}}\qquad%
\subfloat[Fourth layer activation output.]{\label{fig:layer4}\includegraphics[width=0.475\textwidth]{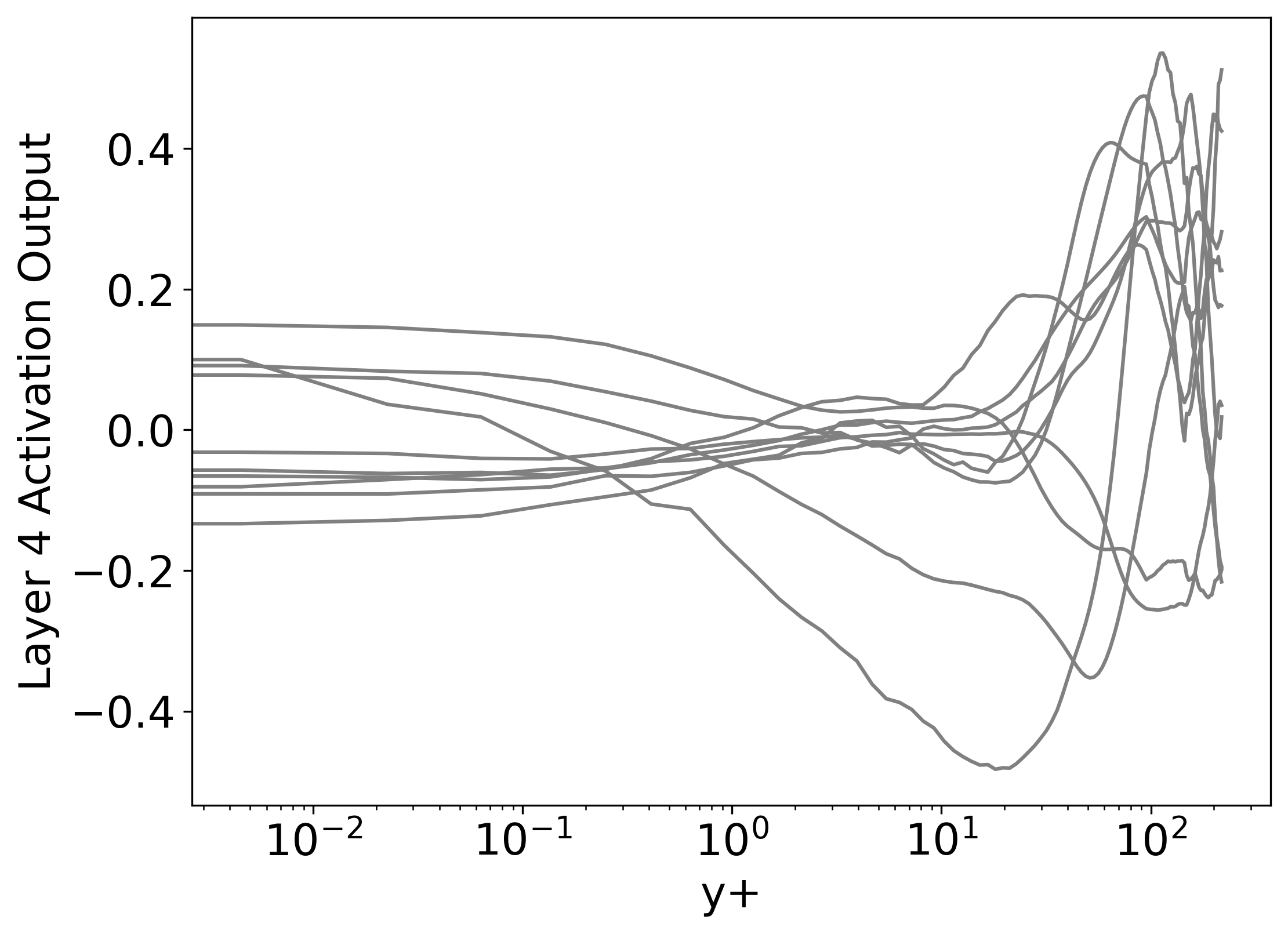}}%
\subfloat[Fifth layer activation output.]{\label{fig:layer5}\includegraphics[width=0.475\textwidth]{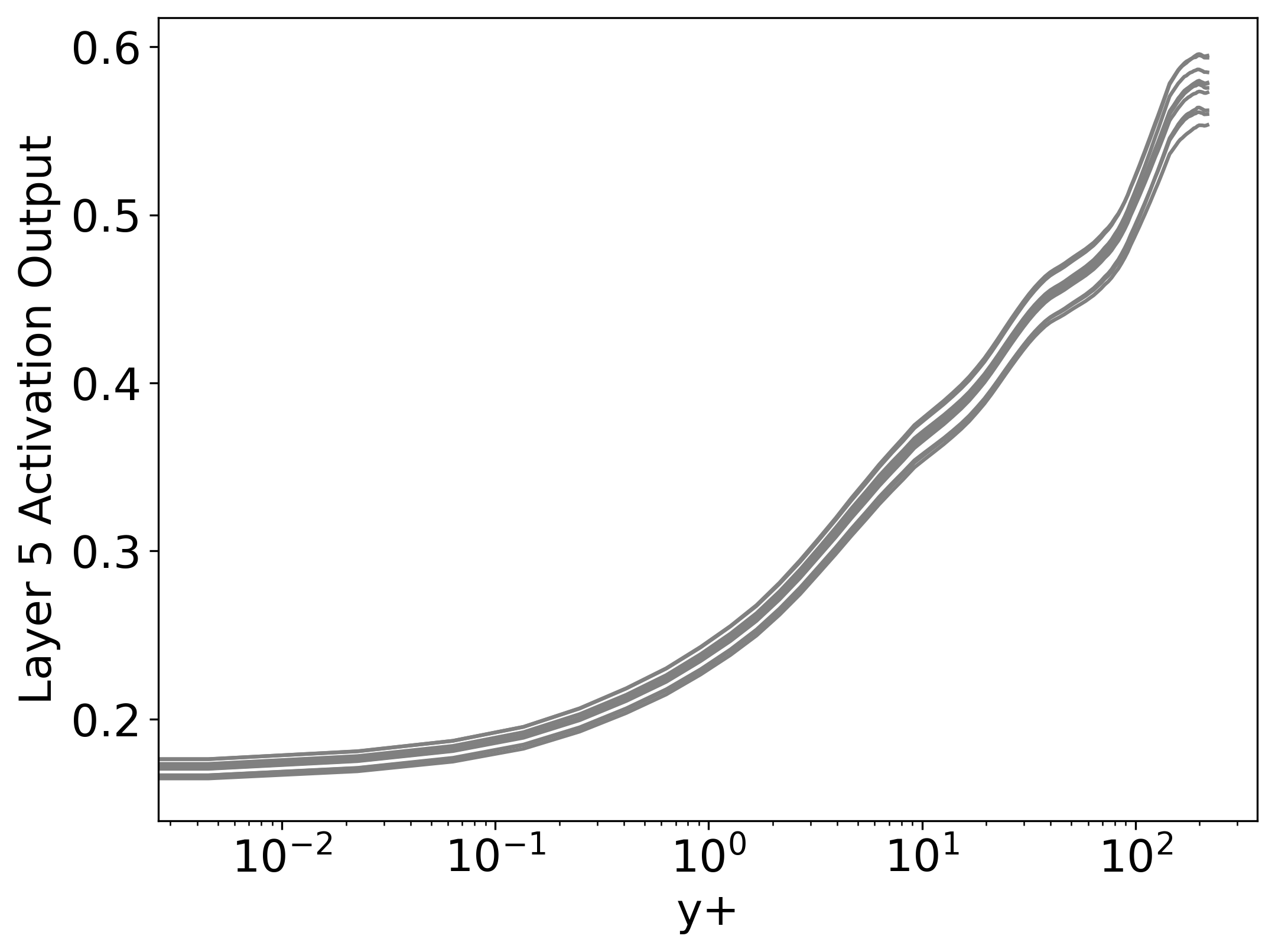}}%
\caption{Activations for turbulent Couette flow Case 2 testing on $Re_{\tau}$=220 with the CNN-BC-$Re_{\tau}$ model. The curves in the plots above represent the activation outputs for each layer. The activation outputs of filters which do not get activated are highlighted with a thicker line.  In (a) one filter is redundant and does not get activated. In (c) two filters and redundant and do not get activated (the activation outputs are horizontal lines). The last layer (e) is also unnecessary as all activation outputs are comparable.}
\label{fig:AV}
\end{figure}

For example, Figure~\ref{fig:AV} shows the activations of different layers of the CNN-BC-$Re_\tau$ model for turbulent Couette flow Case 2. Note that after the last activation, the model would apply a weighted sum operation to the output of the last convolutional layer to obtain the final prediction, and that the trainable weights for the weighted sum operation are initialized at $-\frac{1}{10}$. From the plots, we can see that one filter in layer 1 is redundant and does not get activated. A similar behavior occurs in layer 3 where two filters do not get activated. We further note that the entire last convolutional layer is also unnecessary. By removing these redundancies we can decrease the number of trainable parameters from 10166 to 4856, reduce overfitting, and improve the test $R^{2}$ error from 0.9575 to 0.9936.

As we have shown, this approach can be used to adapt and fine-tune the CNN-BC-Re$_\tau$ model originally developed for turbulent channel flow to different one-dimensional turbulent flows and avoid issues such as overfitting and underfitting. The physics of dissimilar one-dimensional turbulent flows may comprise different levels of complexity which are reflected in the input-output relationship that we are trying to capture between the mean velocity gradients and the anisotropic Reynolds stress tensor. Using this technique we can accommodate our model to several flow configurations and drive our neural network design.

\subsubsection{Occlusion Sensitivity Results}
\label{subsubsec:Occlusion Sensitivity Results}
 
\begin{figure}[htb!]
\centering
\includegraphics[width=0.9\linewidth]{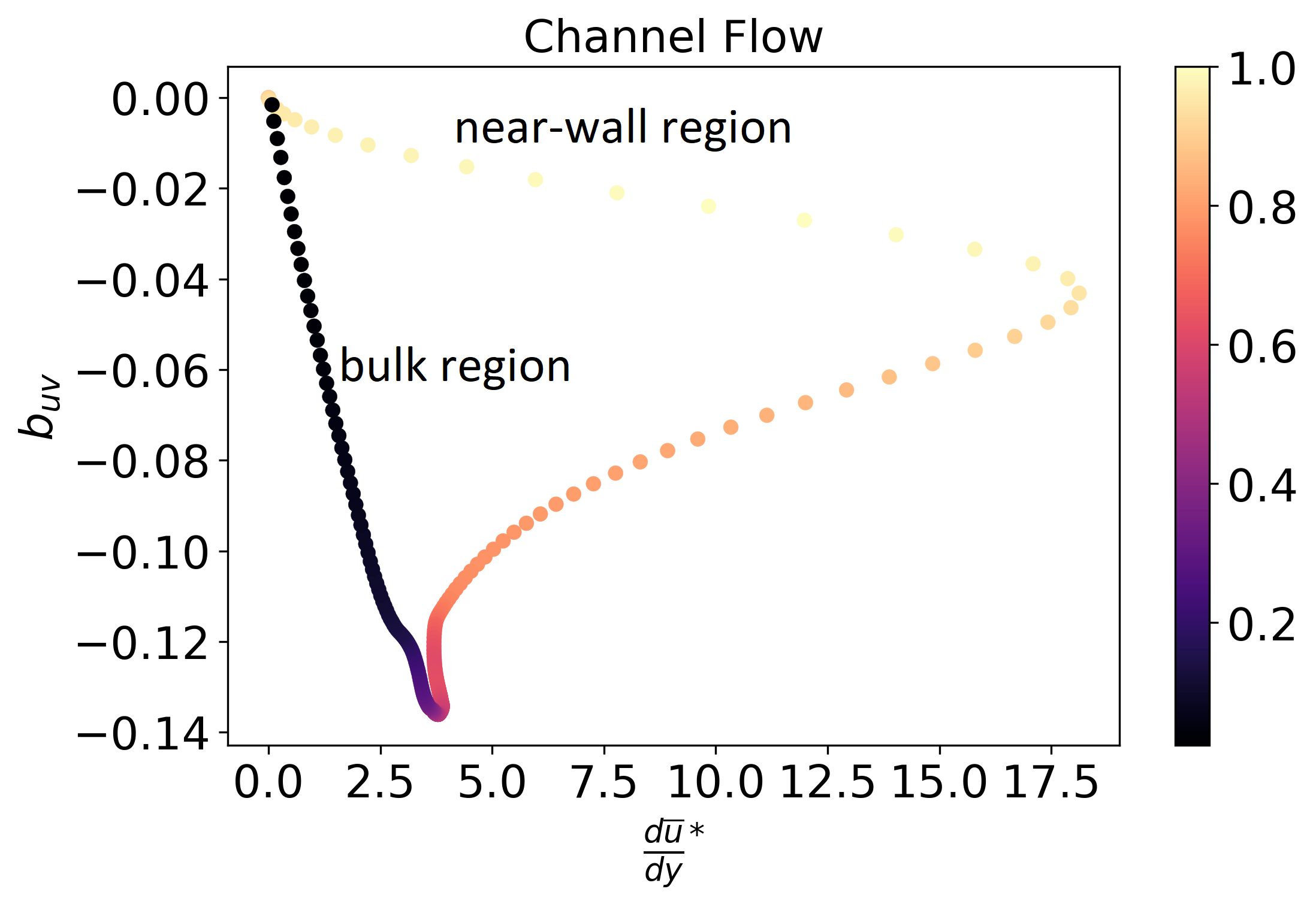}
\caption{Global occlusion sensitivity plot for turbulent channel flow, CNN-BC-$Re_{\tau}$ model for Case 3 testing on friction Reynolds number $Re_{\tau}=1000$. Sliding window covers 50 entries.}
\label{fig:occlusion_1d}
\end{figure}

We start by analysing the global occlusion sensitivity of the models. As discussed in Section~\ref{subsec: Occlusion Sensitivity}, we can determine which regions of the input are important using~\eqref{eq:occlusion_global}. By this, we mean regions which, when occluded, have a greater impact on the overall model performance. To do so, we use a sliding window to systematically occlude different regions of the input. Ideally, the sliding window should be big enough to avoid capturing meaningless local noisy variations but small enough to represent key trends. Given that the number of points in the data change for every friction Reynolds number, the sliding window should be adjusted accordingly. We consistently find the near-wall region to be the most relevant, as expected. Figure \ref{fig:occlusion_1d} shows a global occlusion sensitivity plot for turbulent channel flow. The $x$ and $y$ axes correspond to the input and prediction of the CNN-BC-$Re_{\tau}$ model for Case 3 testing on friction Reynolds number $Re_{\tau}=1000$. The color bar on the right denotes the important regions with light colors. As we can see from the plot, the bulk region on the left is in black, whereas the near-wall region at the top of the plot is colored in light yellow, which shows that the behaviour of the trained model aligns with the underlying physics of the problem.

\begin{figure}[htb!]
\centering
\includegraphics[width=0.8\linewidth]{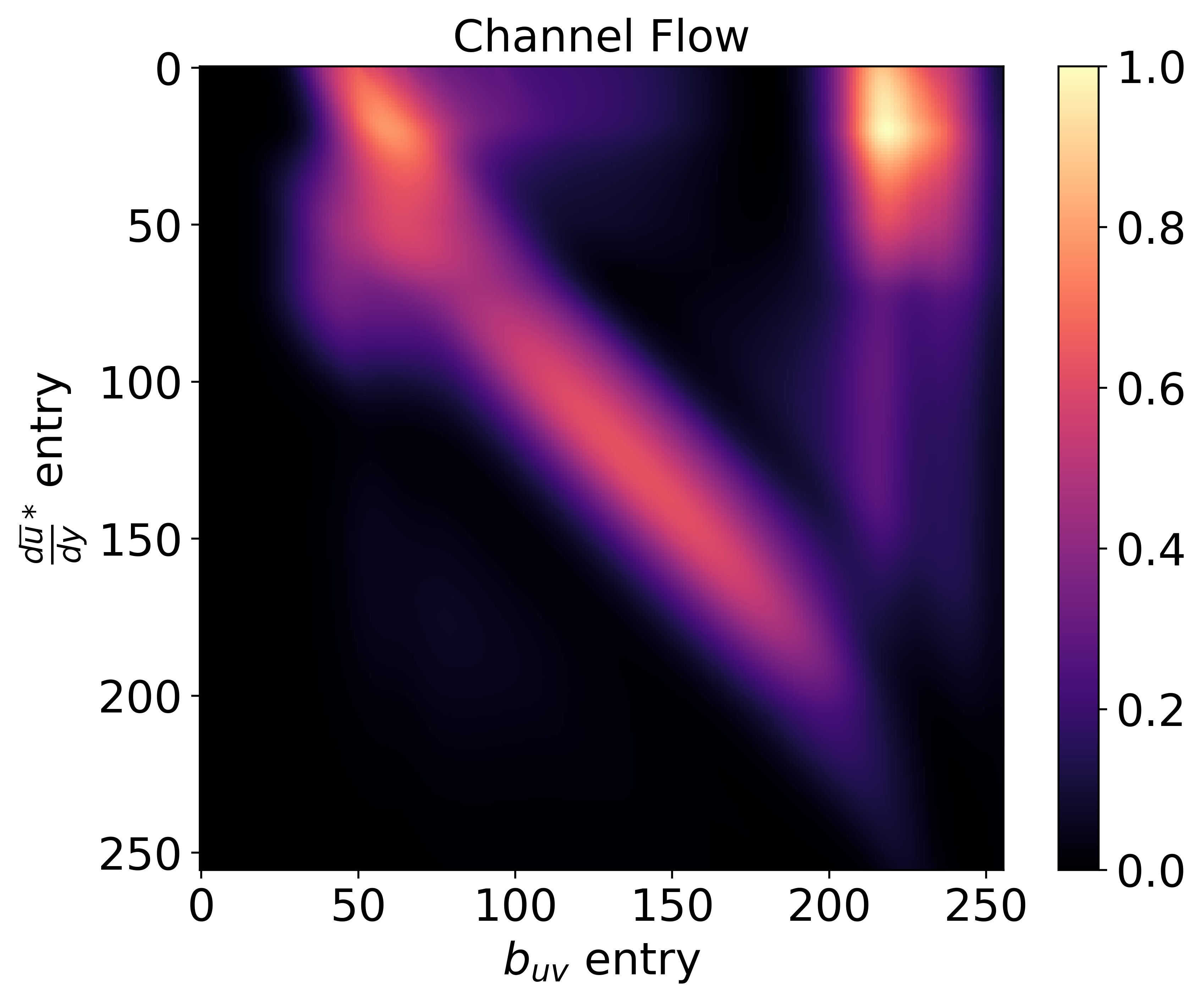}
\caption{Local occlusion sensitivity plot for turbulent channel flow, CNN-BC-$Re_{\tau}$ model for Case 3 testing on friction Reynolds number $Re_{\tau}=1000$. Sliding window covers 50 entries.}
\label{fig:occlusion_2d}
\end{figure}

\begin{figure}[htb!]
\centering
\includegraphics[width=0.8\linewidth]{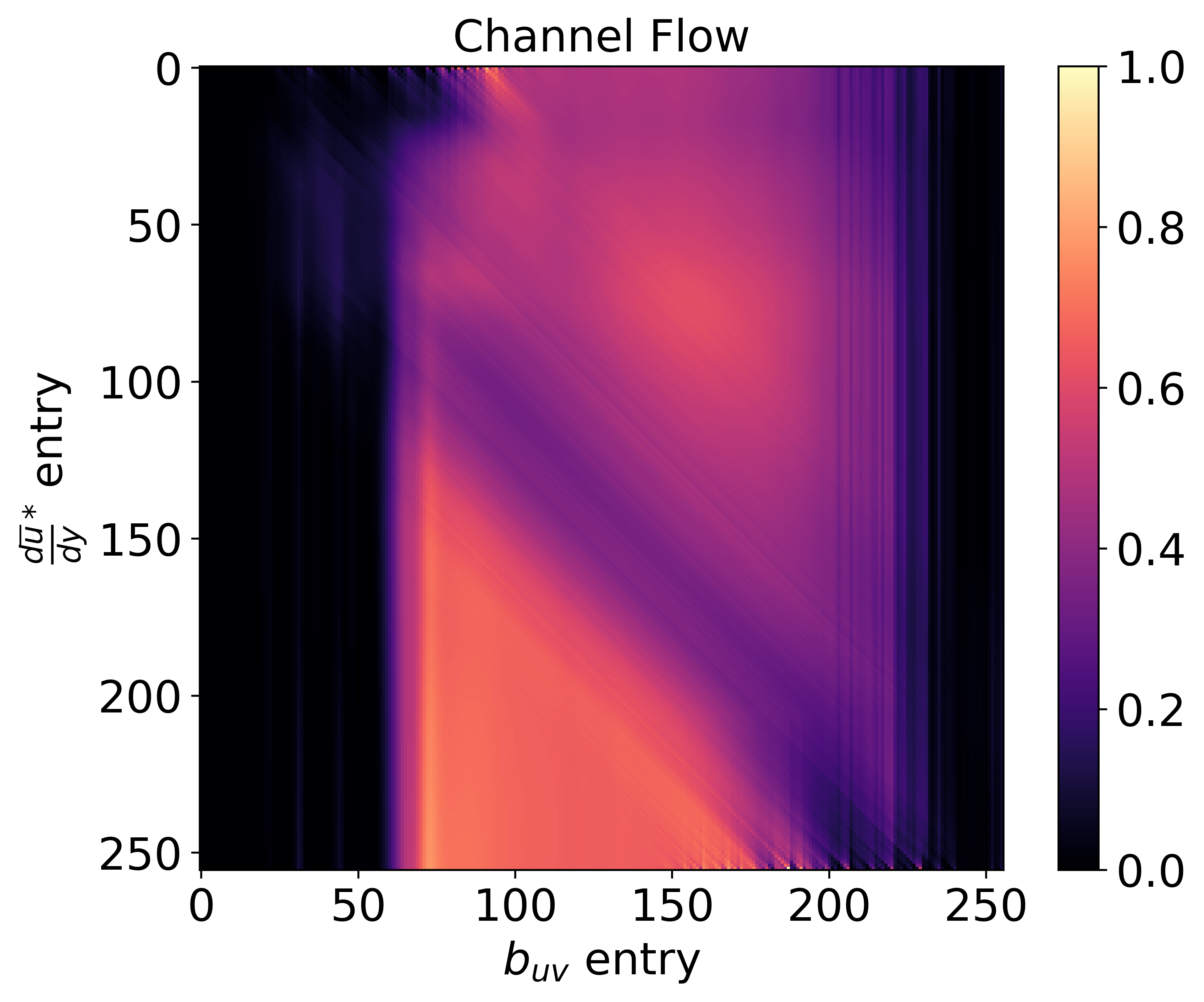}
\caption{Local occlusion sensitivity plot for turbulent channel flow, CNN-BC-$Re_{\tau}$ model for Case 3 testing on friction Reynolds number $Re_{\tau}=1000$ before training.}
\label{fig:occlusion_2d_random_w}
\end{figure}

Figure \ref{fig:occlusion_2d} displays the equivalent local occlusion sensitivity plot for the same model and flow configuration. In this plot, the $y$ axis corresponds to the entry in the input normalized mean velocity gradient array and the $x$ axis to the entry in the output prediction array, so that we can appreciate how occluding different inputs along the height of the channel affect individual $b_{uv}$ predictions. Figure~\ref{fig:occlusion_2d} shows that the model has learnt non-local relationships since the normalized velocity gradient at a given position along the channel height affects the prediction at other locations. For example, the near-wall input entries clearly affect the last few entries of the prediction array as we can see from the yellow patch on the top right corner of the plot.
On the other hand, Figure \ref{fig:occlusion_2d_random_w} shows the local occlusion sensitivity plot for turbulent channel flow before training the CNN-BC-$Re_{\tau}$ model, when the model parameters are still random. The clear differences between Figure \ref{fig:occlusion_2d_random_w} and Figure \ref{fig:occlusion_2d} corroborate that the results are not merely an artefact coming from the interpretability technique itself.

\begin{figure}[htb!]
\centering
\includegraphics[width=0.9\linewidth]{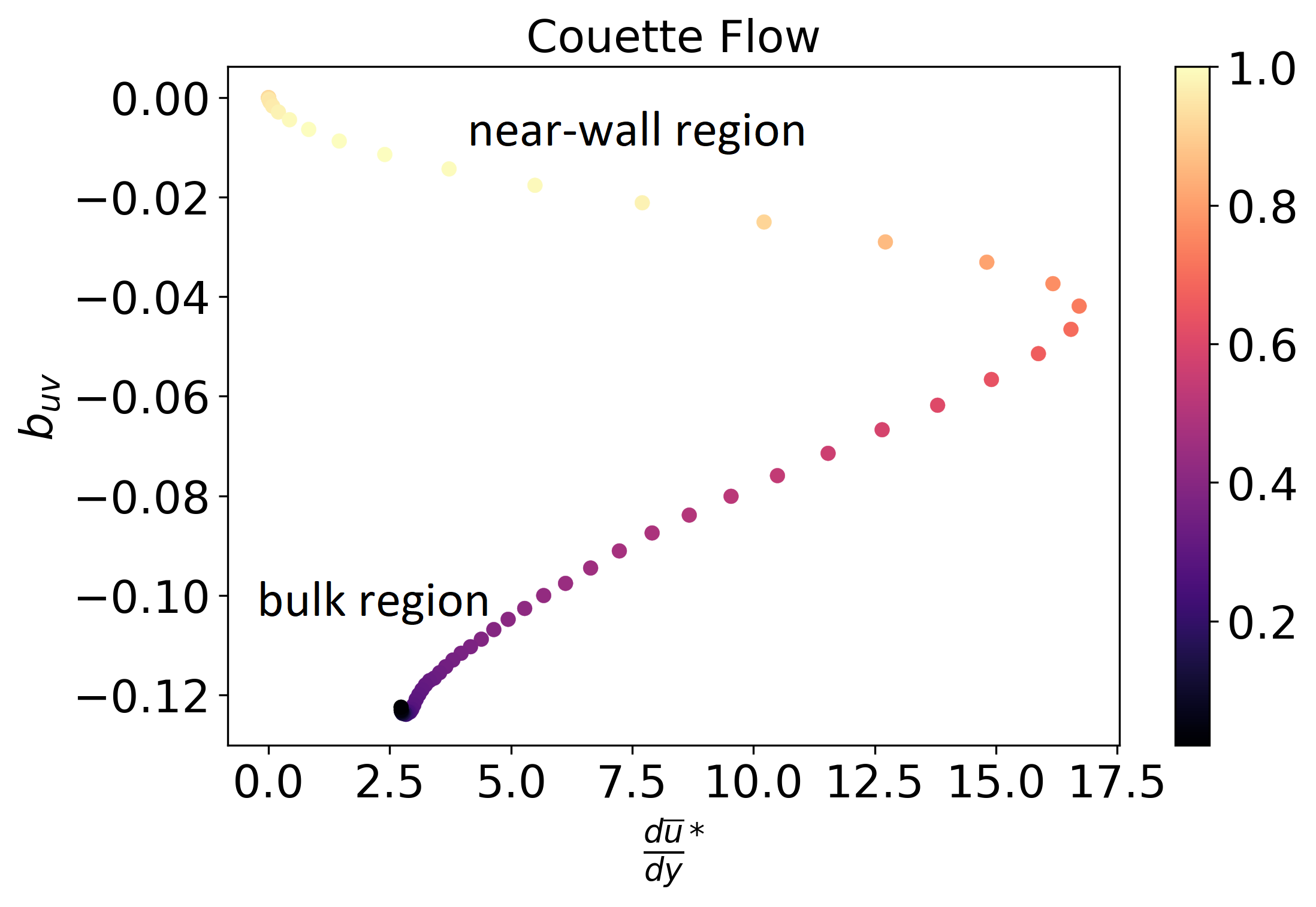}
\caption{Global occlusion sensitivity plot for turbulent Couette flow, CNN-BC-$Re_{\tau}$ model for Case 3 testing on friction Reynolds number $Re_{\tau}=93$. Sliding window covers 50 entries.}
\label{fig:couette_occlusion_1d}
\end{figure}

\begin{figure}[htb!]
\centering
\includegraphics[width=0.8\linewidth]{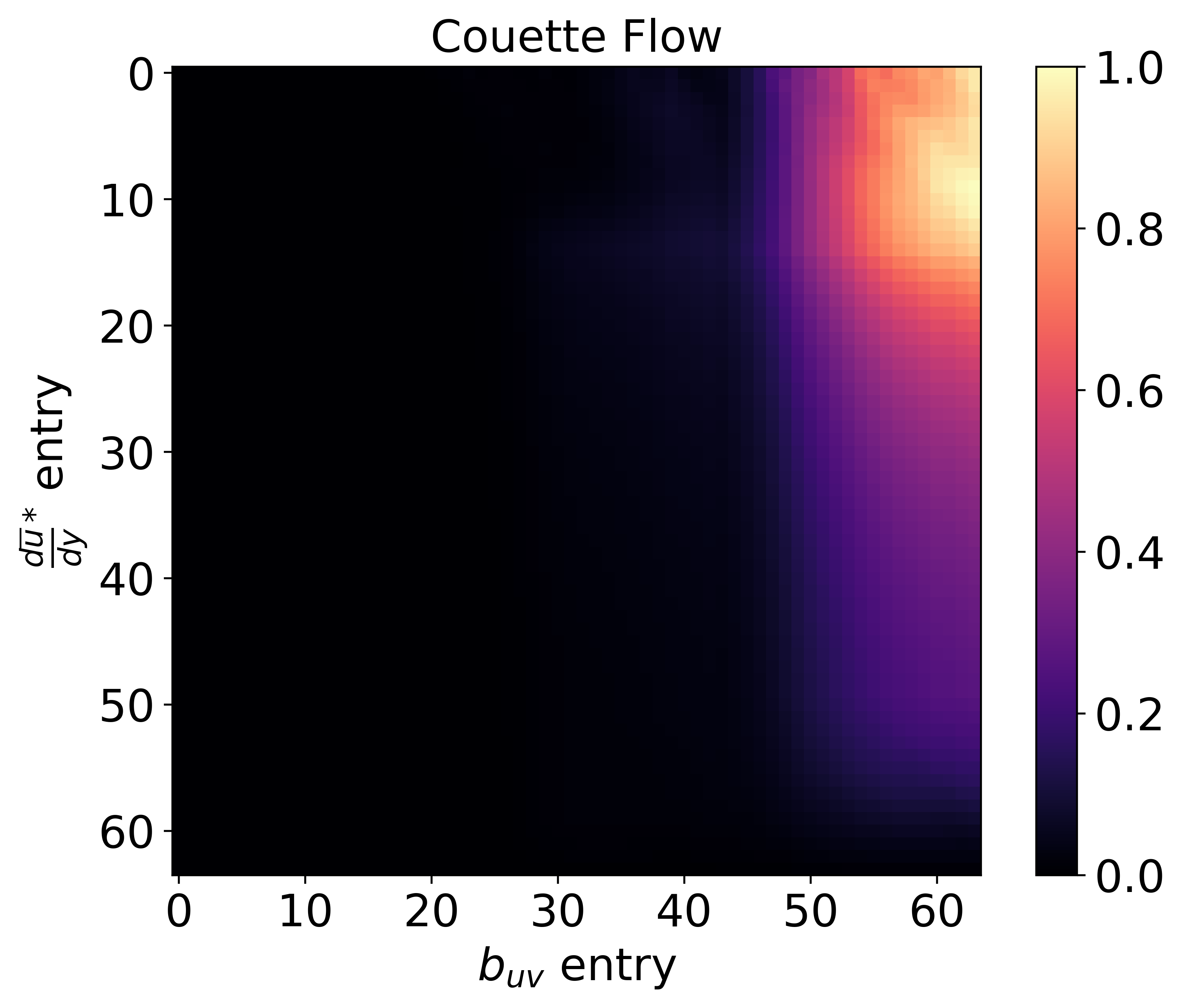}
\caption{Local occlusion sensitivity plot for turbulent Couette flow, CNN-BC-$Re_{\tau}$ model for Case 3 testing on friction Reynolds number $Re_{\tau}=93$. Sliding window covers 50 entries.}
\label{fig:couette_occlusion_2d}
\end{figure}

\begin{figure}[htb!]
\centering
\includegraphics[width=0.9\linewidth]{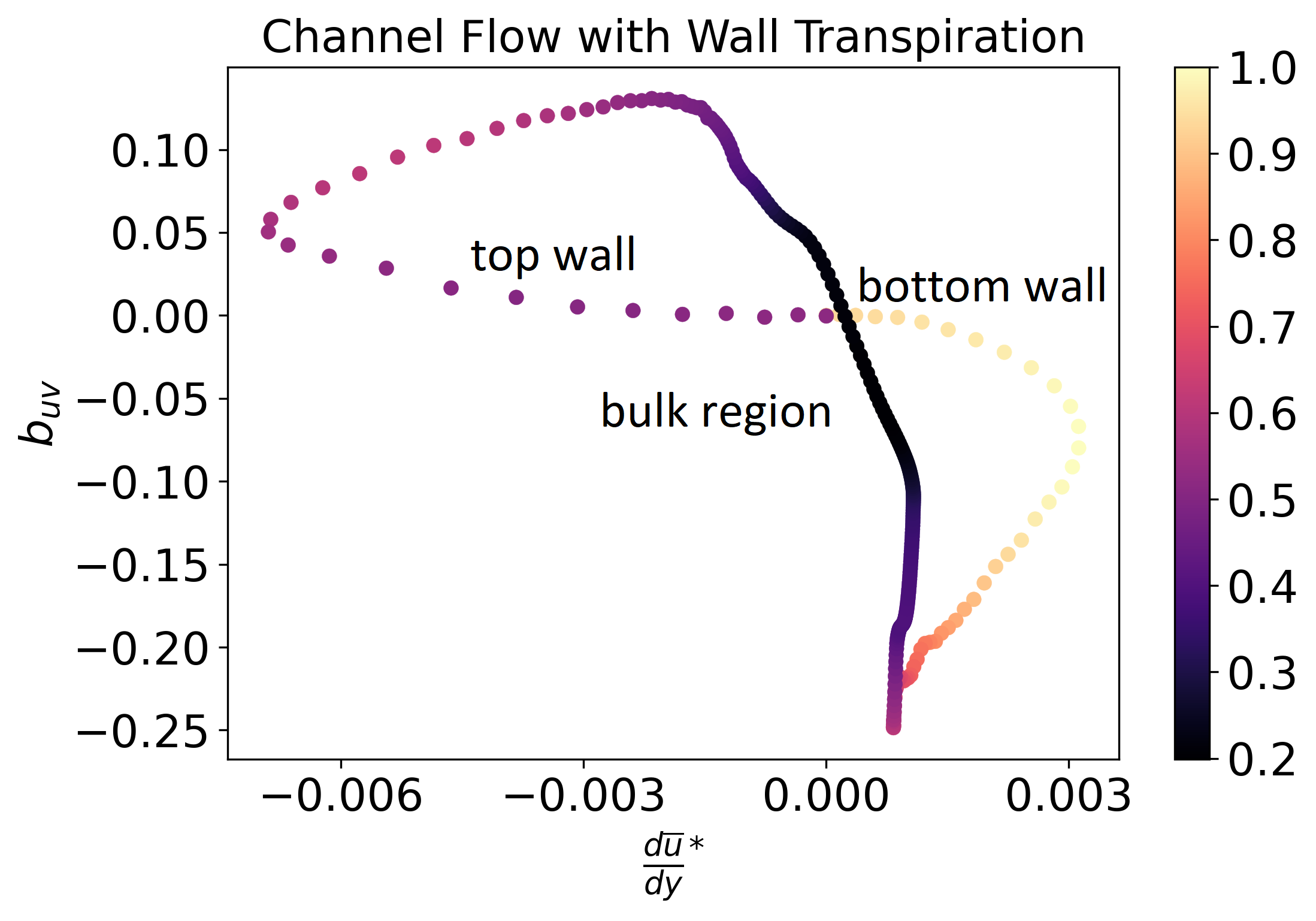}
\caption{Global occlusion sensitivity plot for turbulent channel flow with wall transpiration, CNN-BC-$Re_{\tau}$ model for Case 3 testing on $Re_{\tau}=250$. Sliding window covers 50 entries.}
\label{fig:transpiration_occlusion_1d}
\end{figure}

\begin{figure}[htb!]
\centering
\includegraphics[width=0.8\linewidth]{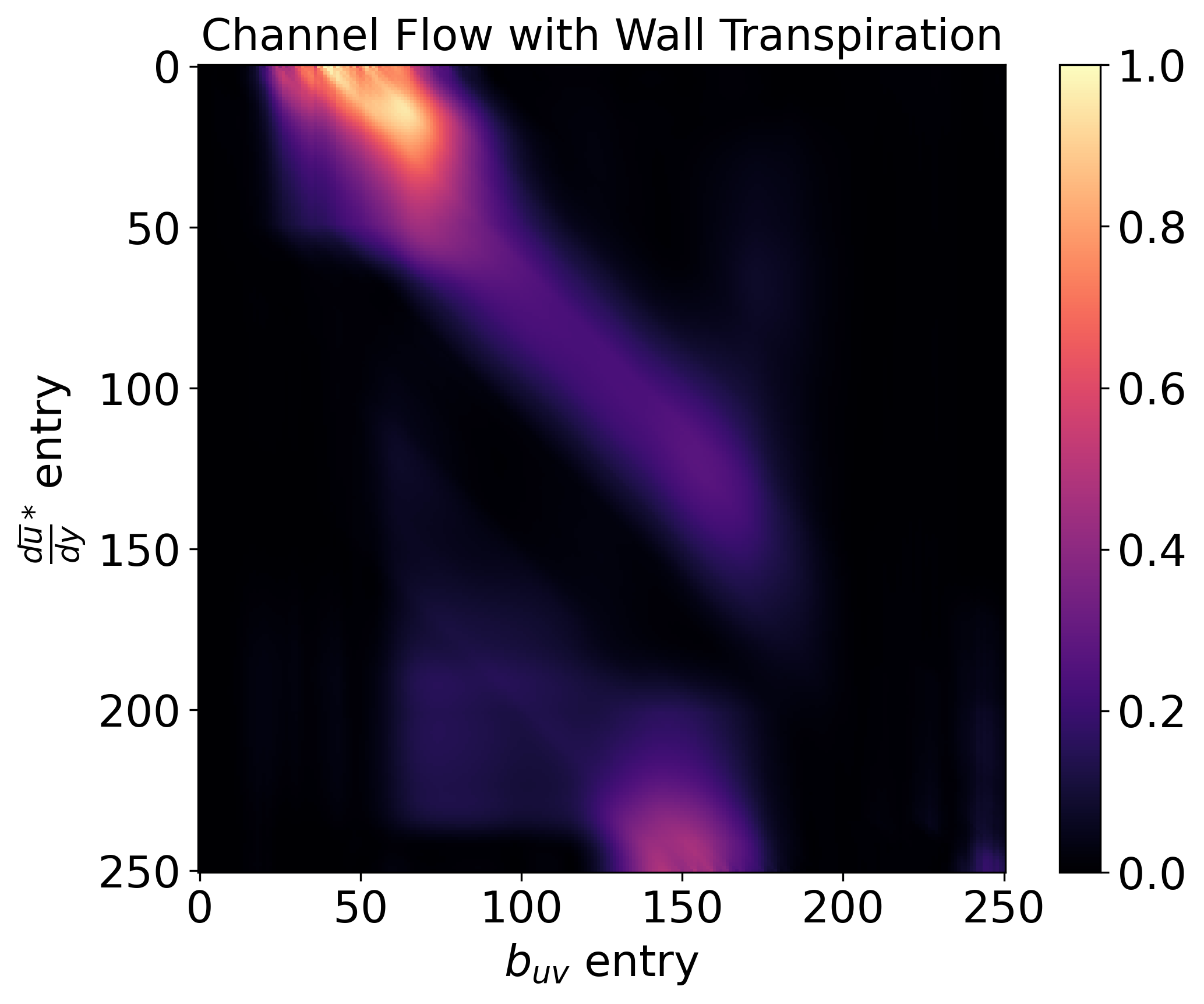}
\caption{Local occlusion sensitivity plot for turbulent channel flow with wall transpiration, CNN-BC-$Re_{\tau}$ model for Case 3 testing on friction Reynolds number $Re_{\tau}=250$. Sliding window covers 50 entries.}
\label{fig:transpiration_occlusion_2d}
\end{figure}

Similar global and local occlusion sensitivity plots can be obtained for turbulent Couette flow and turbulent channel flow with wall transpiration. Figure~\ref{fig:couette_occlusion_1d} and Figure~\ref{fig:couette_occlusion_2d} display sensitivity plots for the CNN-BC-$Re_{\tau}$ model applied to Couette Flow Case 3. The model also pays attention to the near-wall region in this case, as seen in Figure \ref{fig:couette_occlusion_1d}. According to Figure \ref{fig:couette_occlusion_2d}, changes in the first entries of the input array seem to especially affect the last few entries of the output. We indeed found that although the model was obtaining good results near the wall, its predictions for the bulk region tended to fluctuate for similar training losses. We can conclude that the plot reflects that changes in the near wall region input substantially affect the bulk region prediction. Lastly, Figure~\ref{fig:transpiration_occlusion_1d} and Figure~\ref{fig:transpiration_occlusion_2d} represent the occlusion sensitivity of the model for channel flow with wall transpiration. We can see that the model focuses on the near-wall regions where most of the physics takes place and it ignores the bulk, which corresponds to the straight black line near the center of the plot in Figure \ref{fig:transpiration_occlusion_1d}.

\subsubsection{Gradient-based Sensitivity Results}
\label{subsubsec: Gradient-based sensitivity results}

\begin{figure}[htb!]
\centering
\includegraphics[width=0.9\linewidth]{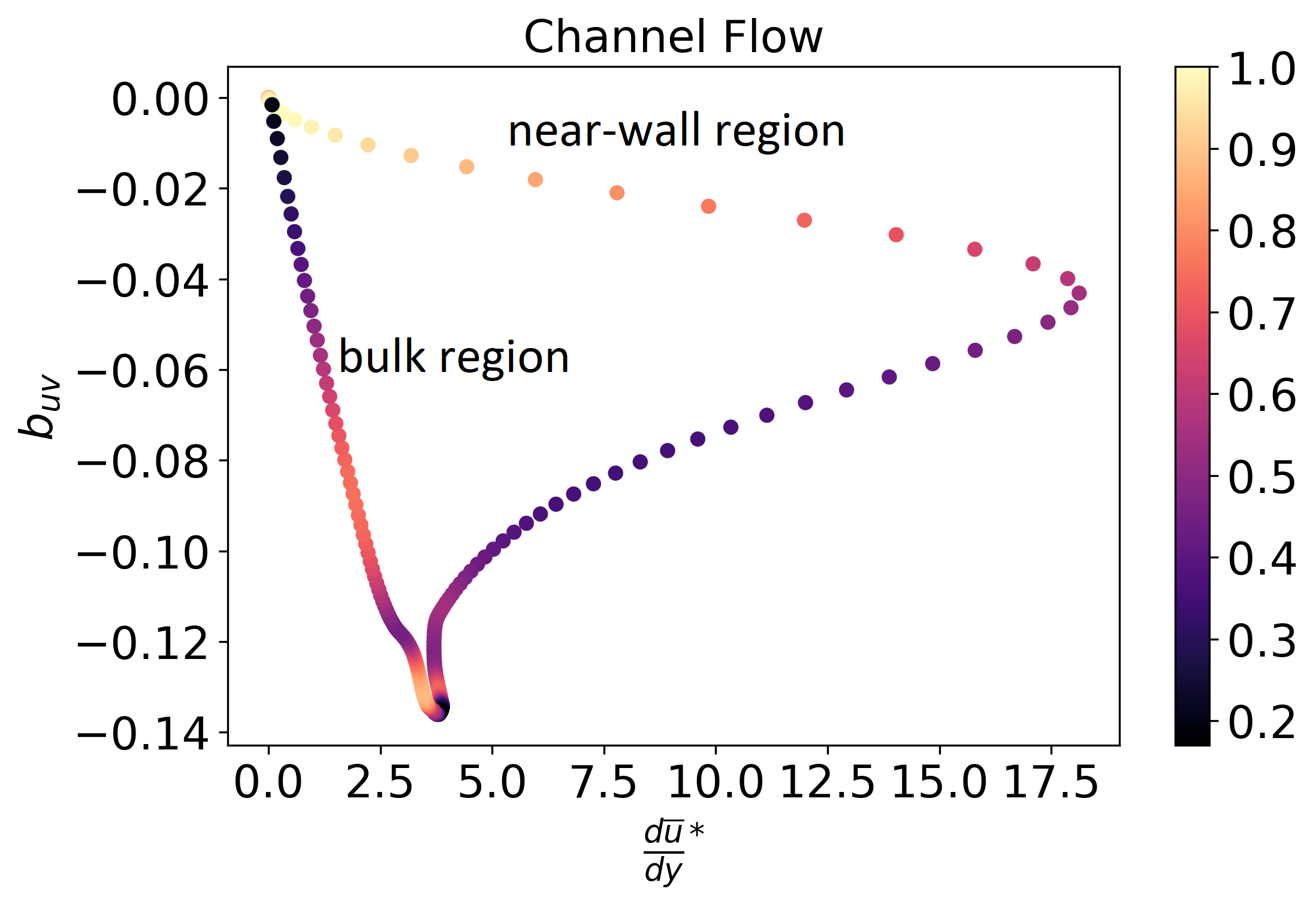}
\caption{Global smooth gradient-based sensitivity plot for turbulent channel flow, CNN-BC-$Re_{\tau}$ model for Case 3 testing on friction Reynolds number $Re_{\tau}=1000$.}
\label{fig:gradients}
\end{figure}

We aim at measuring the relative importance of each entry in the input array on the final model prediction by looking at the magnitude of the gradients with respect to the loss as previously described in Section~\ref{subsec: Gradient-based Sensitivity}. In this context, bigger gradients mean greater importance. Based on~\eqref{eq:grad_global} we can plot smooth global gradient-based saliency maps. Figure~\ref{fig:gradients} displays a plot for the CNN-BC-$Re_{\tau}$ model for turbulent channel flow Case 3 testing on friction Reynolds number $Re_{\tau}=1000$. As in Figure \ref{fig:occlusion_1d}, Figure~\ref{fig:gradients} also highlights the near-wall region. It does, however, slightly highlight other regions. This could be attributed to artifacts coming from the batch normalization layers within the network, which only partially estimate the mean and variance on individual batches and therefore introduce noise. Overall, we find gradient-based and occlusion sensitivity maps tend to agree, but the latter seems to be more consistent and reliable. 

\section{Conclusion}
\label{sec:Conclusion}

The RANS equations are widely used in turbulence modeling applications but accurately modeling the Reynolds stress tensor remains a challenging problem. Recently, researchers have started to explore machine learning applications to turbulence modelling using novel deep neural network architectures trained with DNS data. However, little emphasis has been given to the interpretability of the models. Applying human expertise to tune and improve machine learning models is necessary, especially in such a complex field as turbulence. Understanding the algorithm decision-making process, and how different physics embedding approaches influence it, is key to improving the models.

In this work, we focused on developing better performing neural networks for turbulent one-dimensional flows and applied several interpretability techniques to recognize which inputs have the greatest effect on the accuracy of the model prediction. We compared the results obtained with our CNN models for turbulent channel flow to those using the original FCFF model by~\cite{fang2018deep} and obtained considerable improvements and a good fit to the data. Next, we demonstrated the applicability of our model to other one-dimensional flows, for which it also proved successful. Finally, we proposed a number of interpretability techniques which were used as a guideline to fine-tune the model to different flow configurations and to check that the model behaviour was indeed in line with our understanding of the underlying physics of the problem.

There are several directions for future work, such as testing the models on more turbulent one-dimensional flow configurations and on higher dimensional turbulent flows. The latter will likely require developing new physics embedding techniques as well as more sophisticated interpretability approaches. Furthermore, in some exploratory studies outside the scope of the present work, we also found that applying transfer learning can substantially speed up the training process of the CNN models discussed in this paper. Starting from the model weights for channel flow, the number of epochs required to train the model for Couette flow and channel flow with wall transpiration can be reduced by one order of magnitude compared to retraining the model from scratch. Transfer learning could be explored in future work to help improve the training time of neural networks applied to turbulence modelling.

\clearpage
\bibliography{biblio}
\bibliographystyle{apalike}

\end{document}